\nonstopmode
\documentclass[pdflatex,sn-mathphys-num]{sn-jnl}% Math and Physical Sciences Numbered Reference Style 
\usepackage{graphicx}%
\usepackage{multirow}%
\usepackage{amsmath,amssymb,amsfonts}%
\usepackage{amsthm}%
\usepackage{mathrsfs}%
\usepackage[title]{appendix}%
\usepackage{xcolor}%
\usepackage{textcomp}%
\usepackage{manyfoot}%
\usepackage{booktabs}%
\usepackage{algorithm}%
\usepackage{algorithmicx}%
\usepackage{algpseudocode}%
\usepackage{listings}%
\begin{document}

\title[Article Title]{Real-PGDN: A Two-level Classification Method for Full-Process Recognition of Newly Registered Pornographic and Gambling Domain Names}


\author[1]{\fnm{Hao} \sur{Wang}}\email{hopehaowang@gmail.com}

\author[2]{\fnm{Yingshuo} \sur{Wang}}\email{wangyingshuo@foxmail.com}

\author[3]{\fnm{Junang} \sur{Gan}}\email{ganjunang@gmail.com}

\author*[4]{\fnm{Yanan} \sur{Cheng}}\email{chengyn@hit.edu.cn}

\author[5]{\fnm{Jinshuai} \sur{Zhang}}\email{zhangjs0711@gmail.com}

\affil*[1]{\orgdiv{School of Computer Science and Technology}, \orgname{Harbin Institute of Technology}, \orgaddress{\city{Harbin}, \postcode{150001}, \state{Heilongjiang}, \country{China}}}


\abstract{Online pornography and gambling have consistently posed regulatory challenges for governments, threatening both personal assets and privacy. Therefore, it is imperative to research the classification of the newly registered Pornographic and Gambling Domain Names (PGDN). However, scholarly investigation into this topic is limited. Previous efforts in PGDN classification pursue high accuracy using ideal sample data, while others employ up-to-date data from real-world scenarios but achieve lower classification accuracy. This paper introduces the Real-PGDN method, which accomplishes a complete process of timely and comprehensive real-data crawling, feature extraction with feature-missing tolerance, precise PGDN classification, and assessment of application effects in actual scenarios. Our two-level classifier, which integrates CoSENT (BERT-based), Multilayer Perceptron (MLP), and traditional classification algorithms, achieves a 97.88\% precision. The research process amasses the NRD2024 dataset, which contains continuous detection information over 20 days for 1,500,000 newly registered domain names across 6 directions. Results from our case study demonstrate that this method also maintains a forecast precision of over 70\% for PGDN that are delayed in usage after registration.}




\maketitle

\section{Introduction}\label{sec1}

Domain names are reusable online resources, with hundreds of thousands being registered and reclaimed daily \cite{bib1}. Within these newly registered domain names (NRD), a significant portion is exploited for pornographic and gambling purposes. Websites in these domains pose substantial harm to users across all age groups and often employ deceptive means to capture users' account information. Many Internet users, due to a lack of security awareness, suffer personal information breaches, resulting in potential financial loss. Thus, the timely, efficient, and accurate recognition of PGDN is paramount for safeguarding cyberspace. Scholarly research on this topic is limited. Earlier studies focused on classifying PGDN with the aim of high accuracy using ideal sample data, while other studies utilized recent real-world data but yielded lower classification accuracy\cite{bib2,bib3}. For the recognition of PGDN, both practical applicability and accuracy are crucial.

 In recent years, deep learning has developed rapidly. The classic MLP has gradually played a role in the field of domain name classification, and pre-trained models such as BERT have begun to show their powerful effect in natural language text processing of domain name information. These innovations in deep learning open new possibilities in the domain classification field, which previously relied heavily on traditional classification algorithms. It is worth exploring how to reasonably cooperate with the old and new technologies in the field of PGDN classification and give full play to the best performance.
 
 In addition, the temporal characteristics of domain features are seldom emphasized. New pornographic and gambling domains frequently undergo information changes, such as in DNS and website content, which can be instrumental in domain identification. Moreover, in real-world scenarios, constraints like access restrictions and other objective factors mean the data collected for each domain cannot be complete, necessitating classification methods for PGDN to accommodate missing features.
 
Accordingly, we propose the Real-PGDN method. Through rapid and comprehensive retrieval of real-world domain-related information, coupled with prolonged tracking to capture time-series attributes, we have compiled a dataset, NRD2024, encompassing information on over 1.5 million NRD. We meticulously extract key features for PGDN, annotate 2,000 domains within the NRD2024 dataset, enhance the data, and train the model. We utilize a BERT-based fine-tuning model, CoSENT, to embed text features, concatenating them with MLP outputs to yield the initial binary classification result. The initial classification result is combined with numerical features and processed with a random forest algorithm to obtain the final classification result. Our method attains over 97\% precision and accuracy in recognizing PGDN within real-world data scenarios characterized by missing features. Drawing from the NRD2024 dataset, we annotate 100 domain names with delayed usage for pornographic and gambling purposes and employ the Real-PGDN method for early prediction. The results show that more than 70\% of the domain names are forecasted, and more than 80\% can be detected within 20 days.

The main contributions of this paper are as follows:
\begin{itemize}
\item A new method is proposed for detecting PGDN with high accuracy and feature-missing tolerance in real-world scenarios by employing the two-level classification strategy that integrates CoSENT, MLP, and Random Forest algorithms.

\item A dataset, NRD2024, is obtained, which contains six-dimensional information on over 1.5 million NRD from August 10 to August 16, 2024. We keep track of each domain name for at least 14 days. The dataset is now publicly accessible on Kaggle\footnote{\url{www.kaggle.com/datasets/hopehaowang/nrd2024}}.

\item We conduct a thorough study of the features used for classifying PGDN, recommending a novel set of features suitable for real-world PGDN recognition. This 20-dimensional feature set, derived from six different detecting directions, undergoes secondary processing and selection.
\end{itemize}

\section{The Proposed Real-PGDN Method}\label{sec2}
\begin{figure}[h]
\centering
\includegraphics[width=\textwidth]{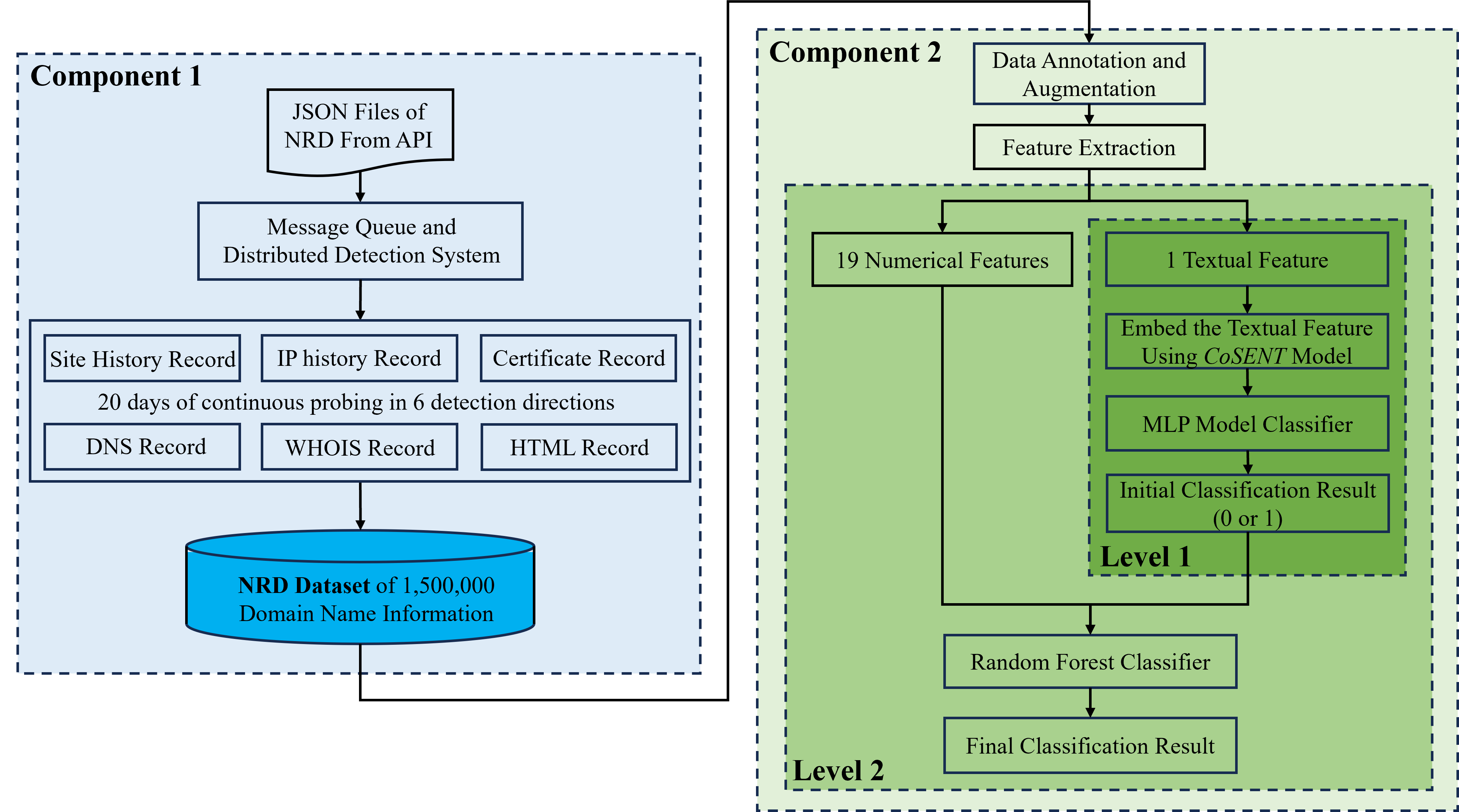}
\caption{\centering Flowchart of the Real-PGDN method.}\label{fig 1}
\end{figure}

A flowchart of the Real-PGDN method is shown in Fig. \ref{fig 1}. As illustrated, the proposed method is segmented into two principal components:

The first component (Section \ref{subsec2 1}) involves the acquisition of raw data from real-life scenarios. Initially, we obtain daily lists of NRD via the domains-monitor website\footnote{domains-monitor.com}. These domain names are then inserted into a RabbitMQ message queue. Subsequently, each domain name undergoes 6 detecting tasks within our detection system's nodes: Site History Records, IP History Records, Certificate Records, DNS Records, WHOIS Records, and HTML Records. After 20 days of continuous detection, we compile the NRD2024 dataset consisting of data on 1.5 million domain names (Section \ref{subsec2 2}).

The second component encompasses feature engineering and the construction and training of models (Section \ref{subsec2 3}). Initially, we conduct an in-depth analysis of features specific to PGDN, identifying 20 key features from the NRD2024 dataset across four feature sets, which include 19 numerical features and one textual feature. The subsequent step involves data annotation and augmentation. We manually annotate 2,000 domain names for training and validation and respectively augment some of the data using deletion and replacement strategies. Then, we use our two-level classification schema to build a PGDN classifier. We employ the CoSENT model for textual feature embedding and utilize a Multilayer Perceptron to complete the first level of classification. We blend the Level 1 result with the 19 numerical features and apply traditional classification algorithms for secondary classification to obtain the Level 2 result. In Section \ref{subsec3 1}, we will evaluate the adaptability of various traditional classification methods to our model architecture.

\subsection{Data Collection}\label{subsec2 1}

\subsubsection{Comprehensiveness requirements}\label{subsubsec2 1 1}

Our objective is to gather the most comprehensive information on NRD to facilitate PGDN feature analysis, thereby identifying the most effective feature combinations. Algorithm \ref{algo 1} shows the pseudocode of the detection process.

We perform a detection every 24 hours across the following dimensions:(1) DNS Records: We perform standard DNS queries using programmatic interfaces with default parameters, covering all common types such as A, AAAA, NS, MX, CNAME, SOA, and TXT for each domain name. (2) IP History Records: Based on DNS-A records (IPv4 addresses) obtained from prior queries, we utilize a third-party platform API\footnote{api.uouin.com} to compile a list of domain names previously associated with these IP addresses. Given that IP addresses mapped to domain names can change over time, we dynamically update this list throughout the detecting period. (3) HTML Records: By issuing GET requests via HTTPS and HTTP protocols, we gather the complete HTML content of accessible sites associated with each domain, extracting pivotal information such as titles, keywords, and descriptions. (4) Certificate Records: For domains using the HTTPS protocol, we acquire their digital certificate information. Through SSL/TLS protocol wrapping of TCP connections, we extract the digital certificate offered by the target server during the handshake, ultimately parsing it into the X.509 format. Once the certificate is initially detected, detection of this dimension is ceased.	

For domain information that remains static over brief periods, we ensure a timely acquisition within our detection cycle: (1) WHOIS Records: We query WHOIS servers via our detection server cluster. (2) Site History Records: Generally, domain usage tends to have continuity. Domains with a history of abuse are less likely to be selected by reputable users in new lifecycles, leading to lower prices. Conversely, domains used for malicious activities have shorter lifetimes, necessitating frequent domain changes, making low-cost domains ideal. Hence, we collect some site history information from a third-party platform\footnote{seo.juziseo.com}.

\begin{algorithm}
\caption{Detection Process of Domain Name Related Information}\label{algo 1}
\begin{algorithmic}[1]
\While{every 24h}
    \For{each $domain$ in $domains$}
        \For{each $i$ in DNS record types}
            \State $DNSresults[i] \gets$ query DNS records of this domain
            \If{$i$ is DNS-A \& $DNSresults[i]$}
                \State $IP2DomainList \gets$ do IP reverse lookup
            \EndIf
        \EndFor
        \State $htmlPage \gets$ send GET Request to this domain
        \If{this site uses HTTPS \& $Flag = false$}
            \State $certificate \gets$ get certificate of this domain
            \If{$certificate \neq null$}
                \State $Flag \gets true$
            \EndIf
        \EndIf
    \EndFor
\EndWhile
\For{each $domain$ in $domains$}
    \For{each $j$ in other detection directions}
        \State $results[j] \gets$ conduct specific detection
    \EndFor
\EndFor
\end{algorithmic}
\end{algorithm}

\subsubsection{Speed requirements}\label{subsubsec2 1 2}

Given our study's necessity to daily acquire each type of information described in Section \ref{subsubsec2 1 1} for millions of domains, a simple serial detecting model is inadequate for speed requirements. Thus, we devised a distributed detection system leveraging RabbitMQ message queues \cite{bib4}.

\begin{figure}[h]
\centering
\includegraphics[width=1.0\textwidth]{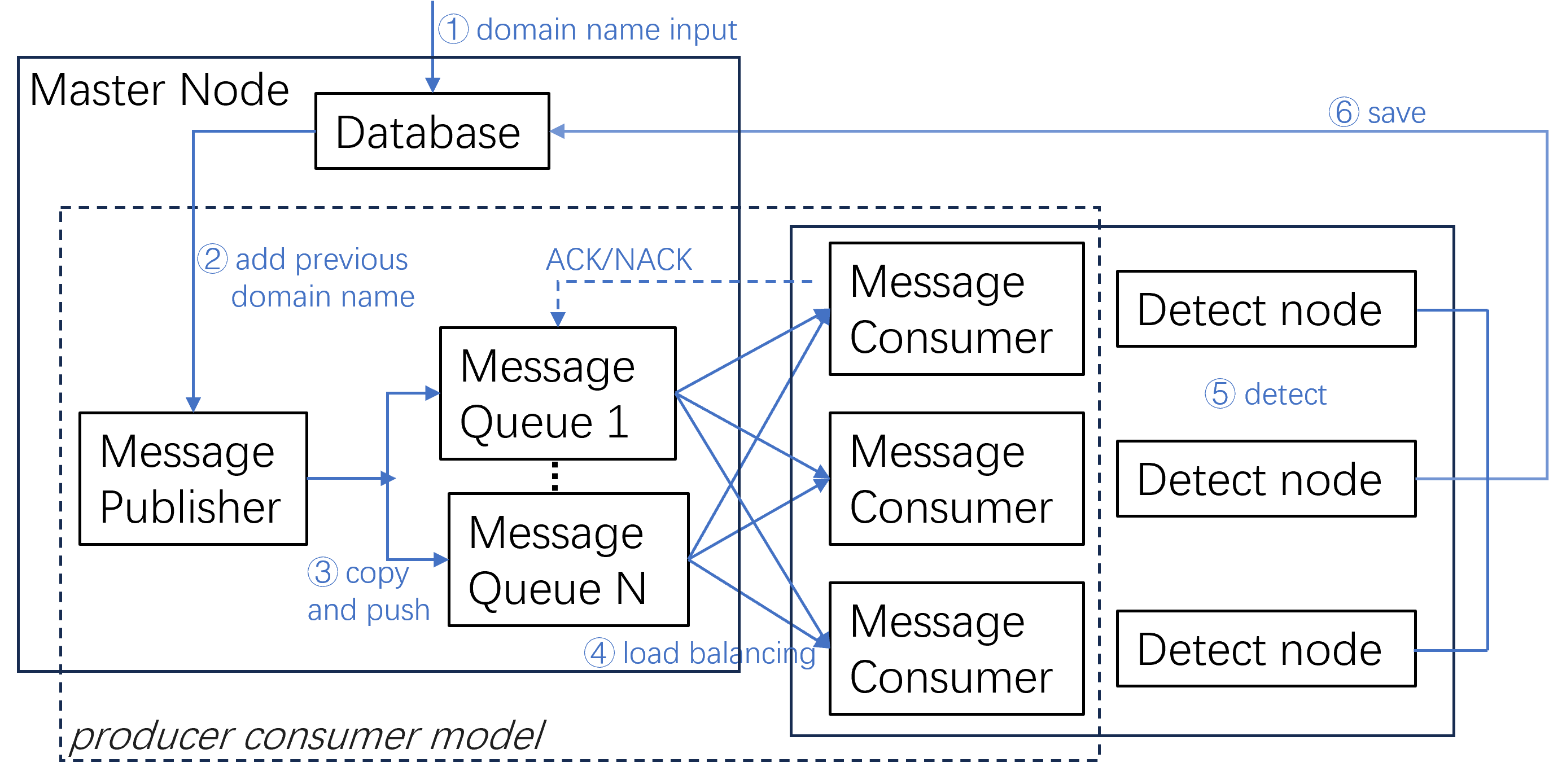}
\caption{\centering Architecture of the distributed detection system.} 
\label{fig 2}
\end{figure}

This system consists of four hosts: one as the main node and the others as subordinate nodes. All nodes function as detection nodes, utilizing RabbitMQ’s producer-consumer model \cite{bib5} for load balancing. Specifically, the main node hosts a MySQL database and RabbitMQ message queue server, acting as the publisher of pending domain messages and the repository for detection results. Subordinate nodes interact with the main node through RabbitMQ and MySQL remote connections. The system architecture is illustrated in Fig. \ref{fig 2}. First, the main node uses the API to acquire new domain data. Second, data is submitted to the main node's RabbitMQ server, where a producer program standardizes the raw domain data. Third, standardized domain data is duplicated across multiple message queues, with each queue containing a complete set of domains for detecting identical information types. Fourth, detection nodes competitively retrieve data from the queues via consumer programs using RabbitMQ. Upon receiving data, each detection node independently detects information relevant to the domain. Fifth, each detection node individually completes the information detection. Finally, results are periodically committed to the MySQL database.

\subsection{NRD2024 Dataset}\label{subsec2 2}

\begin{figure}[h]
\centering
\includegraphics[width=0.55\textwidth]{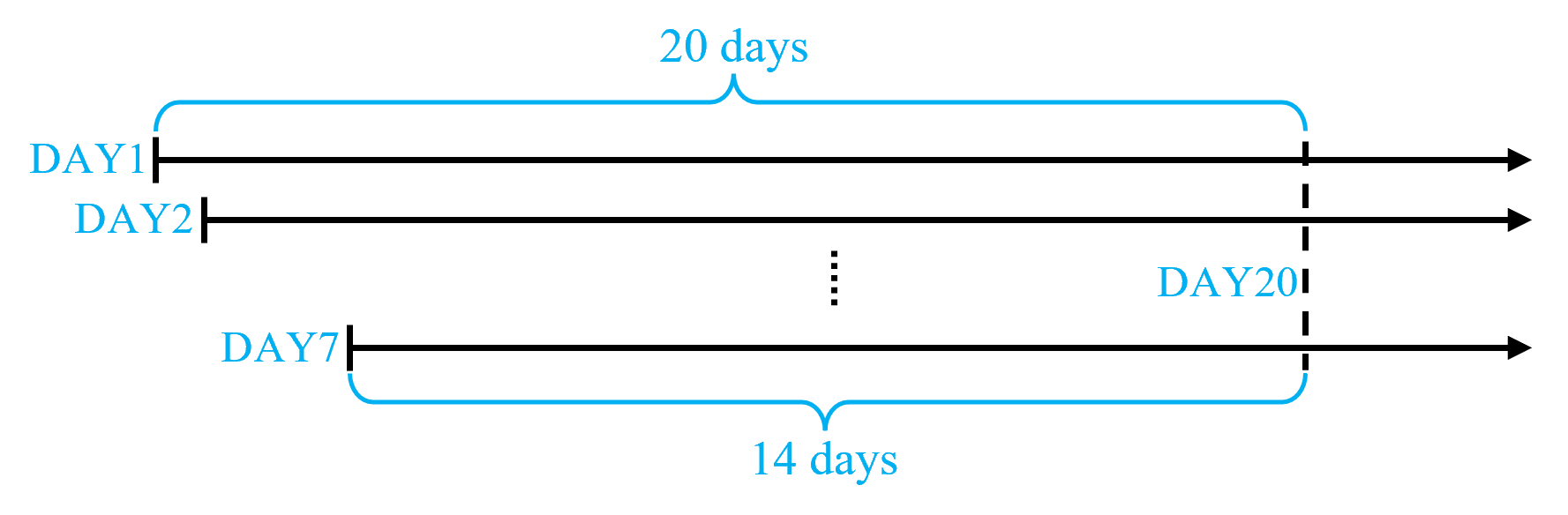}
\caption{\centering Detection timeline.}\label{fig 3}
\end{figure}

Utilizing the detection system introduced in Section \ref{subsec2 1}, we independently completed the data collection for 1,522,363 NRD between August 10 and August 16, 2024, as illustrated in Table \ref{tab 1}. For each domain, we conduct detection every 24 hours starting from the day of registration and continuing for 20 days. This ensures that each NRD has at least 14 days of continuous detection data, with domains registered on the first day undergoing 20 days of detection. Fig. \ref{fig 3} depicts the timeline. We ultimately gather over 10 million pieces of information. We also analyze the composition of top-level domains (TLD) within our dataset, revealing that the \textit{.com} domain, as the largest TLD globally, comprises over 50\% of the registrations over these 7 days, with other popular TLDs each accounting for less than 5\% (See Table \ref{tab 2}).

\begin{table}[h]
\caption{\centering Newly registered domain names over a week.}\label{tab 1}%
\begin{tabular}{@{\hspace{40pt}}c@{\hspace{40pt}} c@{\hspace{40pt}}}
\toprule
\textbf{Date}       & \textbf{Number of Domain Names} \\ 
\midrule
2024-08-10 & 284192                 \\ 
2024-08-11 & 233559                 \\ 
2024-08-12 & 189098                 \\ 
2024-08-13 & 160788                 \\ 
2024-08-14 & 189426                 \\ 
2024-08-15 & 233122                 \\ 
2024-08-16 & 232178                 \\ 
\textbf{Total} & \textbf{1522363}      \\ 
\botrule
\end{tabular}
\end{table}

The data we collected are categorized into six record types, each with detailed field names and descriptions, as shown in Table \ref{tab 3}.

\begin{table}[h]
\caption{\centering Field information of NRD2024 Dataset.}\label{tab 3}%
\begin{tabular}{@{\hspace{10pt}}p{2cm}@{\hspace{10pt}} p{3cm}@{\hspace{10pt}} p{6.5cm}@{\hspace{10pt}}}
\toprule
 \textbf{Type}  &  \textbf{Details}  &  \textbf{Description}  \\ 
\midrule
DNS            & A                                         & IPv4 address record                                              \\ 
                       & AAAA                                      & IPv6 address record                                              \\ 
                       & CNAME                                     & Canonical name record, which specifies alias names               \\ 
                       & MX                                        & Mail exchange record, which routes requests to mail servers      \\ 
                       & NS                                        & Name server record, which delegates a DNS zone to an authoritative server \\ 
                       & SOA    & Start of authority record, which specifies authoritative information about a DNS zone \\ 
                       & TXT                                       & Human-readable notes or machine-readable data of the domain name \\ 
\midrule
 HTML           & Title                                     & Title of the website                                            \\ 
                       & Keywords                                   & Keywords of the website                                        \\ 
                       & Description                                & Description of the website                                     \\ 
                       & Raw HTML                                   & Source code of the website                                     \\ 
\midrule
Certificate        & Certificate Issuer                        & The issuer of the certificate                                   \\ 
                       & Certificate Duration                      & The period of validity of the certificate                       \\ 
\midrule
 WHOIS          & Registration Duration                     & Duration of this domain name registration                       \\ 
                       & Registrar                                  & A company, person, or office that manages the reservation of Internet domain names \\ 
\midrule
IP History      & Domain Name List                          & A list of domain names used by this IP                         \\ 
\midrule
 Site History     & Age                                       & The age of the website                                          \\ 
                       & Historical Title List                     & A list of historical titles used by this site                  \\ 
\botrule
\end{tabular}
\end{table}
\begin{table}[h]
\caption{\centering Top TLDs in new domain name registration.}\label{tab 2}%
\begin{tabular}{@{\hspace{40pt}}c@{\hspace{40pt}} c@{\hspace{40pt}}}
\toprule
\textbf{TLD}       & \textbf{Number of Domain Names} \\ 
\midrule
com       & 778280                 \\ 
shop      & 59225                  \\ 
top       & 55432                  \\ 
xyz       & 53037                  \\ 
online    & 50063                  \\ 
org       & 44943                  \\ 
bond      & 40550                  \\ 
net       & 40437                  \\ 
uk        & 32497                  \\ 
OTHER     & 367899                 \\ 
\textbf{Total} & \textbf{1522363}      \\
\botrule
\end{tabular}
\end{table}

\subsection{Feature Analysis of PGDN}\label{subsec2 3}

This task is based on the data collected in the NRD2024 dataset. We aim to identify features that assist in PGDN classification, which we have divided into four different feature sets, as illustrated in Table \ref{tab 4}. We employ a mixture of features newly utilized in the classification of PGDN and some previously used in related fields. The four feature sets are described as follows.

\begin{table}[h]
\caption{\centering Real-PGDN Features}\label{tab 4}%
\begin{tabular}{@{\hspace{8pt}}p{2.5cm}@{\hspace{8pt}} p{8cm}@{\hspace{8pt}} p{1cm}@{\hspace{10pt}}}
\toprule
\textbf{FeatureSet}      & \textbf{FeatureName}                         & \textbf{Type}                           \\ 
\midrule
\multirow{4}{*}{\centering Site Features} & Title                                       & string                                   \\ 
                        & IP URL redirection metric                   & float                                    \\ 
                        & Enumeration of Certificate Issuers          & int                                      \\ 
                        & Certificate Duration                        & int                                      \\ 

\multirow{8}{*}{\centering DNS Features}  & Oscillation Metric of DNS-A                 & float                                   \\ 
                        & Oscillation Metric of DNS-AAAA              & float                                   \\ 
                        & Oscillation Metric of DNS-NS                & int                                     \\ 
                        & The number of domains with the same DNS-CNAME value & float                              \\ 
                        & DNS-SOA TTL                                 & int                                      \\ 
                        & DNS-SOA Refresh Time                         & int                                      \\ 
                        & DNS-SOA Retry Time                          & int                                      \\ 
                        & Whether the domain name has DNS-MX record   & bool                                     \\ 

\multirow{6}{*}{\shortstack{Domain Name \\ Features}} & Enumeration of TLD                          & int                                      \\ 
                        & Number of vowels in SLD                     & int                                      \\ 
                        & Number of digits in SLD                     & int                                      \\ 
                        & Whether the domain name uses IDN            & bool                                     \\ 
                        & Enumeration of Domain Name Registrars       & int                                      \\ 
                        & Domain name registration duration             & int                                      \\ 

\multirow{2}{*}{\centering History Features}          & Total years of website establishment         & int                                      \\ 
                        & The number of domain names that the IP has used & int                                  \\ 
\botrule
\end{tabular}
\end{table}

\subsubsection{Site Features}\label{subsubsec2 3 1}

\begin{figure}[h]
\centering
\includegraphics[width=0.8\textwidth]{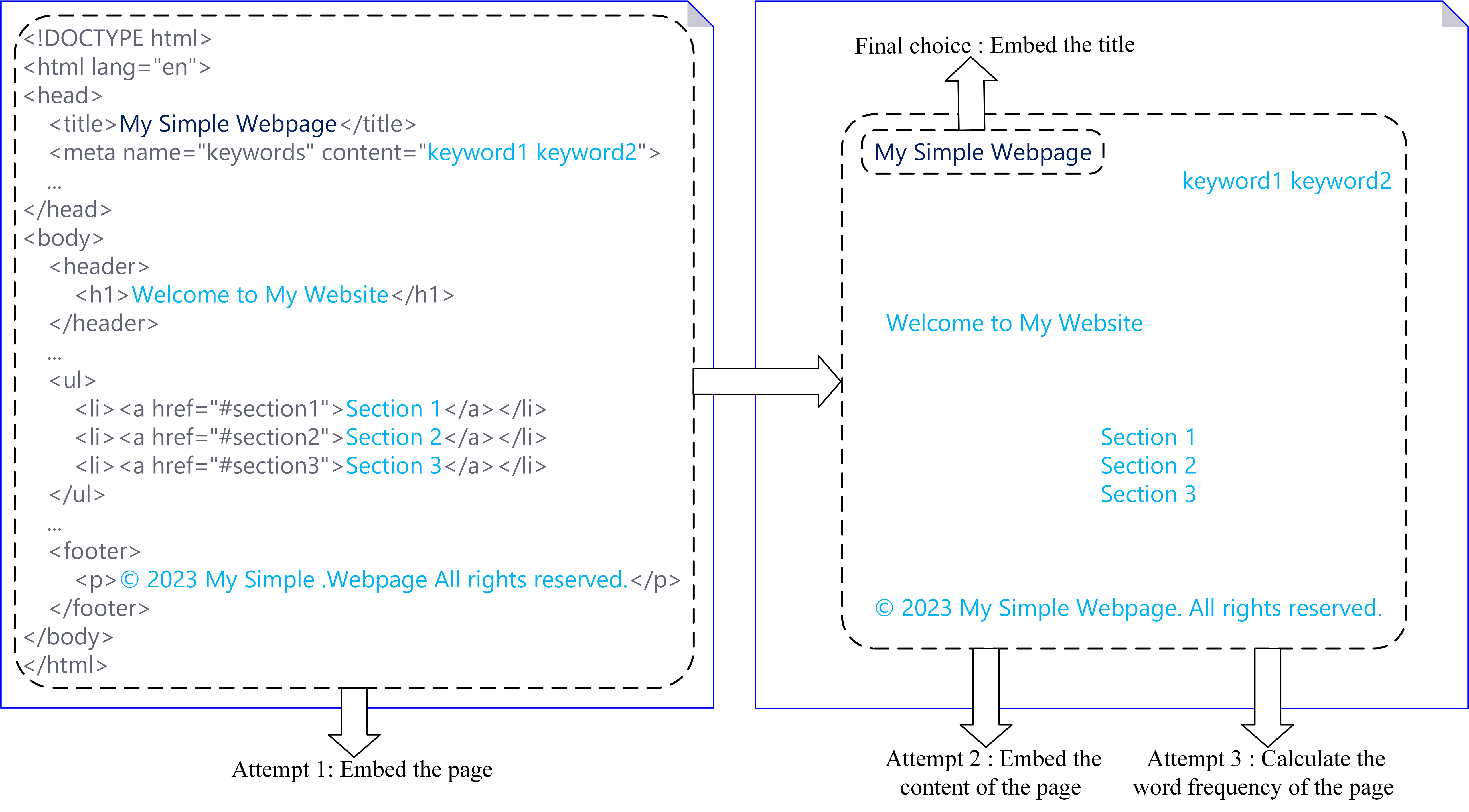}
\caption{\centering Attempts to extract web page content features.}\label{fig 4}
\end{figure}

\textbf{Site Content Feature} Why do we only use titles? We indeed collected the complete HTML file, and the following attempts were made (as shown in Fig. \ref{fig 4}):

\begin{enumerate}[1.]

\item Embed the entire HTML file.

\item Extract the natural language parts from all HTML tags to avoid disruption caused by tags and delimiters and embed them.

\item Extract the natural language parts from all HTML tags, further calculate their respective word frequencies, and do statistical analysis.
\end{enumerate}

However, these attempts are found to be minimally helpful for classification, even detrimental. We speculate this might be due to the ambiguous thematic content in HTML's natural language text, which typically contains a large amount of similar, generic text. Ultimately, we opt to use only the content of the \textit{title} tag, as it usually provides the most concise summary of the page content and shows significant variation between pages. Subsequently, the issue arises with missing \textit{title} tags, which will be addressed in Section \ref{subsec2 5}, detailing the text embedding techniques used for titles and strategies for handling title missing.

\begin{figure}[h]
\centering
\includegraphics[width=0.4\textwidth]{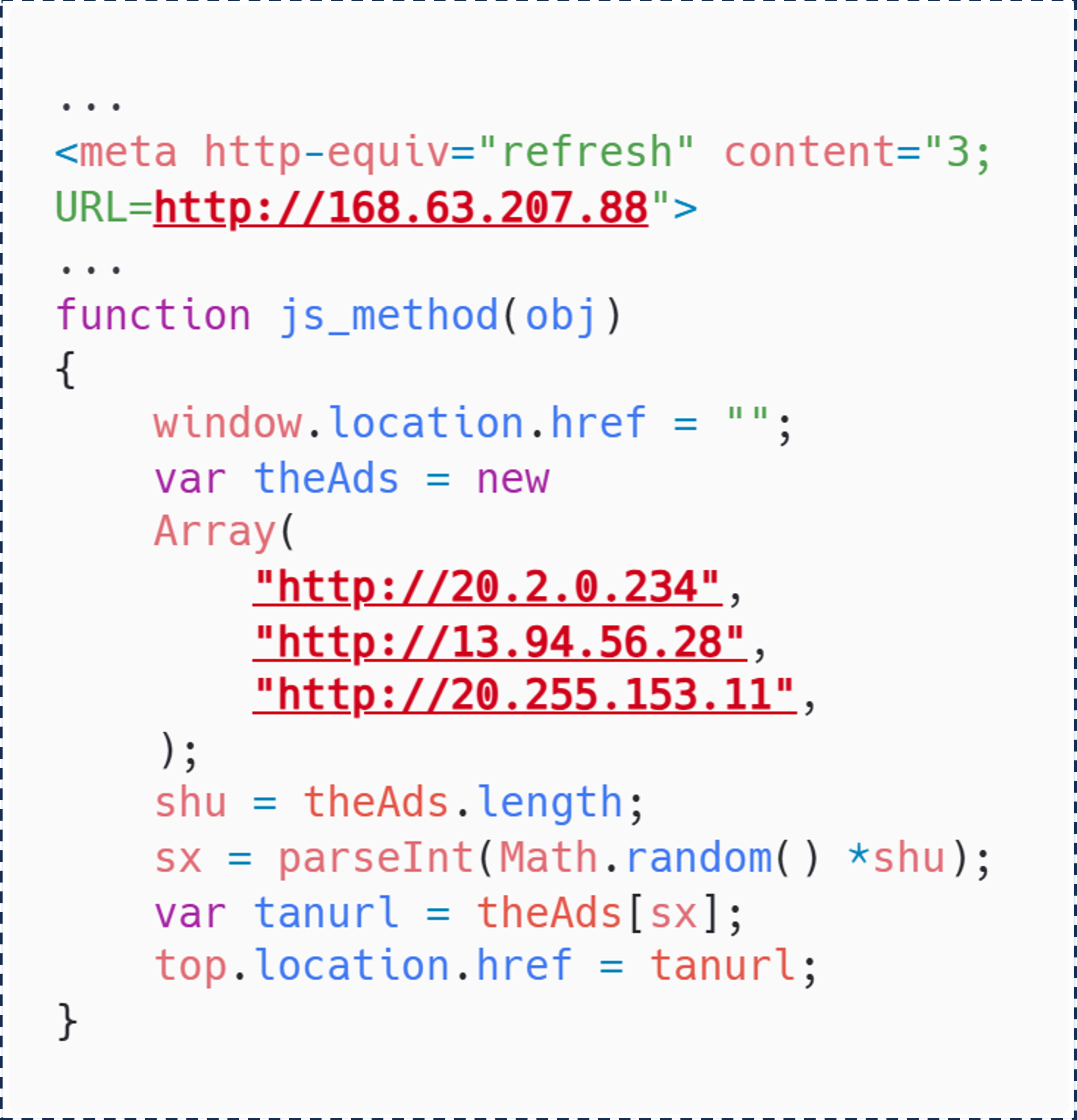}
\caption{\centering An example of the IP URL redirection pages.}\label{fig 5}
\end{figure}

To enhance the identification of redirect pages serving pornographic and gambling sites, we Introduce a feature concerning redirects. First, we count the number of redirect URLs in the HTML file that do not resolve to domain names(as shown in Fig. \ref{fig 5}) using regular expressions, then calculate the ratio \textit{k} of this count \textit{m} to the total length of the HTML file string \textit{l}. We refer to \textit{k} as \textit{IP URL redirection metric} (as shown in Equation \ref{eq 1})

\begin{equation}
{k}=\frac{m}{l}=\frac{ \text { IP URL Redirection Count }}{\text { Total Length of HTML }}
\label{eq 1}
\end{equation}

\begin{figure}[h]
\centering
\includegraphics[width=0.6\textwidth]{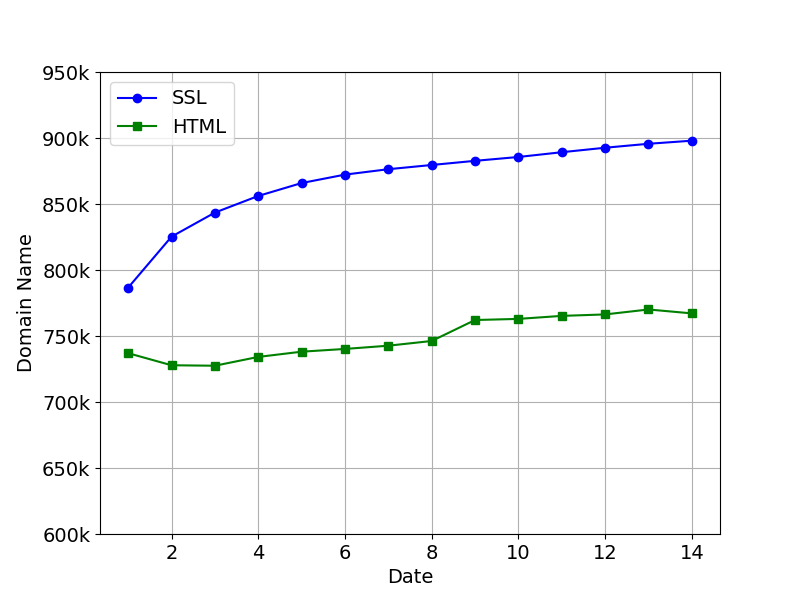}
\caption{Statistics of 1.5 million domain names successfully accessing pages and establishing TLS/SSL connections within 14 days after registration.}\label{fig 6}
\end{figure}

\textbf{Site Certificate Features} A site is considered fully constructed only when we can receive a 200 Status Code via HTTP/HTTPS GET requests. On the other hand, digital certificates are returned by the server side during the TLS/SSL handshake when establishing a client-server connection. We note a gap between the domain registration and the completion of site construction during which we can establish a TLS/SSL connection but cannot access meaningful site content. The difference between the two curves in Fig. \ref{fig 6} highlights this gap. Therefore, we consider extracting TLS/SSL certificates from sites that do not yield meaningful web content, allowing us to analyze them before the sites become fully operational. We extract two features: the certificate issuer and the duration of the certificate.

\subsubsection{DNS Features}\label{subsubsec2 3 2}

In this section, we introduce the use of time-series features of DNS-A/AAAA and DNS-NS in the classification of PGDN for the first time. We also pay attention to two often overlooked DNS records in previous research: DNS-CNAME and DNS-SOA.

\textbf{Effective Utilization of DNS Time Series} We have discovered that pornographic and gambling domains, along with other malicious domains, tend to ``oscillate" between a limited set of IP addresses to evade detection. We use our proposed \textit{Oscillating Metric} (as shown in Equation \ref{eq 2}) to quantify this phenomenon in the DNS time series. Taking IP addresses as an example, for a given domain name, we calculate the sum \textit{c} of changes in IP from day \textit{i} to day \textit{i-1} using DNS-A/AAAA records, as well as the total number of different IPs over \textit{n} days, and then compute the ratio of \textit{c} to \textit{s}. We applied this strategy to both DNS-A/AAAA and DNS-NS features.

\begin{equation}
  Oscillating\ Metric =\frac{c}{s}=\frac{\sum_{i=1}^{n} \sum_{j=1}^{m} 1_{\left\{\text {data }_{i j} \neq S e t_{i-1}\right\}}}{\sum_{i=1}^{n} \sum_{j=1}^{m} 1_{\left\{\text {data }_{i j} \neq {\bigcup_{k=1}^{i}}Set_{k}\right\}}}
\label{eq 2}
\end{equation}

\textbf{DNS-CNAME and DNS-SOA Reords} CNAME records facilitate the mapping between different domain names and are commonly used for domain aliases. When multiple domains are set with the same CNAME record, changing the DNS record of the target domain can also alter the DNS records of all domains referencing it. In our detection cycle, NRD possessing CNAME records are rare; statistically, less than 5\% of domains are detected with at least one CNAME record. Among these, pornographic and gambling domains constitute the majority, leading us to believe that CNAME records facilitate frequent changes in DNS information for these domains.

\begin{figure}[h]
\centering
\includegraphics[width=0.6\textwidth]{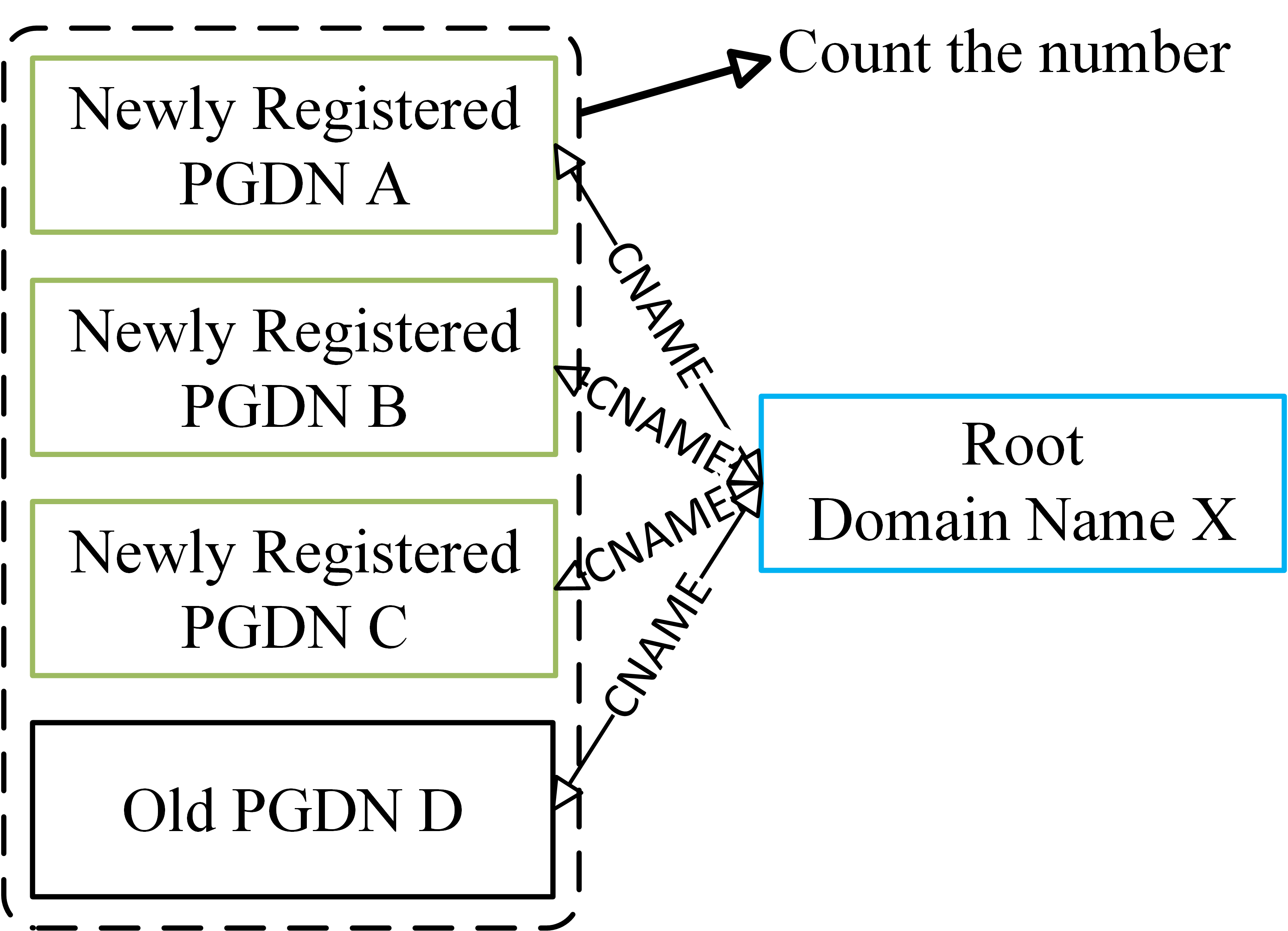}
\caption{\centering Many-to-one mapping of CNAME records for newly registered domain names.}\label{fig 7}
\end{figure}

To prevent domains that legitimately use CNAME services from being misclassified, we perform some filtering on domain CNAME records. We determine whether a domain has a CNAME record, and for those that do, we conduct reverse queries on the referenced domains and tally the number of records(as shown in Fig. \ref{fig 7}).

To further delineate the characteristics of the NS servers used by the domains, we also collect DNS-SOA records. The DNS-SOA record is rarely focused on in previous research. It contains key information about the DNS server, such as the administrator's email address, the refresh time frame of the server, domain update history, etc. We extract the DNS zone's refresh time, retry time, and the TTL values of the record cache as features.

\subsubsection{Domain Name Features and History Features}\label{subsubsec2 3 3}

For domain features, we draw on feature extraction methods used in previous research \cite{bib6,bib7,bib8}. Domains are divided into TLDs and SLDs, with TLDs being finite and enumerated as features. Since SLDs are often random and lack semantic clarity, we do not apply NLP techniques to SLDs. Instead, we extract character count features of SLDs. Additionally, we employ some common WHOIS features such as registrars and registration duration.

We consider the history features of domains to complement the three main types of features previously mentioned. The historical usage of a domain determines its reputation, which significantly influences the domain's future usage. We document the earliest construction records of the domains and use the total years of construction as a feature. We also calculate the number of different IPs used by the domain during the detection period as a second feature, reflecting the historical reputation of the IPs utilized by the domain.

\subsection{Data Annotation}\label{subsec2 4}

We cannot find a blocklist focused exclusively on newly registered PGDN, and most existing institutional data do not cover all such domains. In fact, a significant number of these domains are never tracked or managed. Hence, we do not use data provided by third-party institutions or government agencies. Instead, we opt for random sampling of domains by our experts who then manually annotate them. We set the labels for \textit{Pornography}, \textit{Gambling}, and \textit{Others}. Our annotation relies entirely on the real and real-time web content of the evaluated datasets, i.e. the title and raw HTML parts extracted on the day of crawling. Furthermore, we assign the target page's label to redirect pages, a task that has to be performed by manually accessing the redirect web pages.

\begin{figure}[h]
\centering
\includegraphics[width=0.95\textwidth]{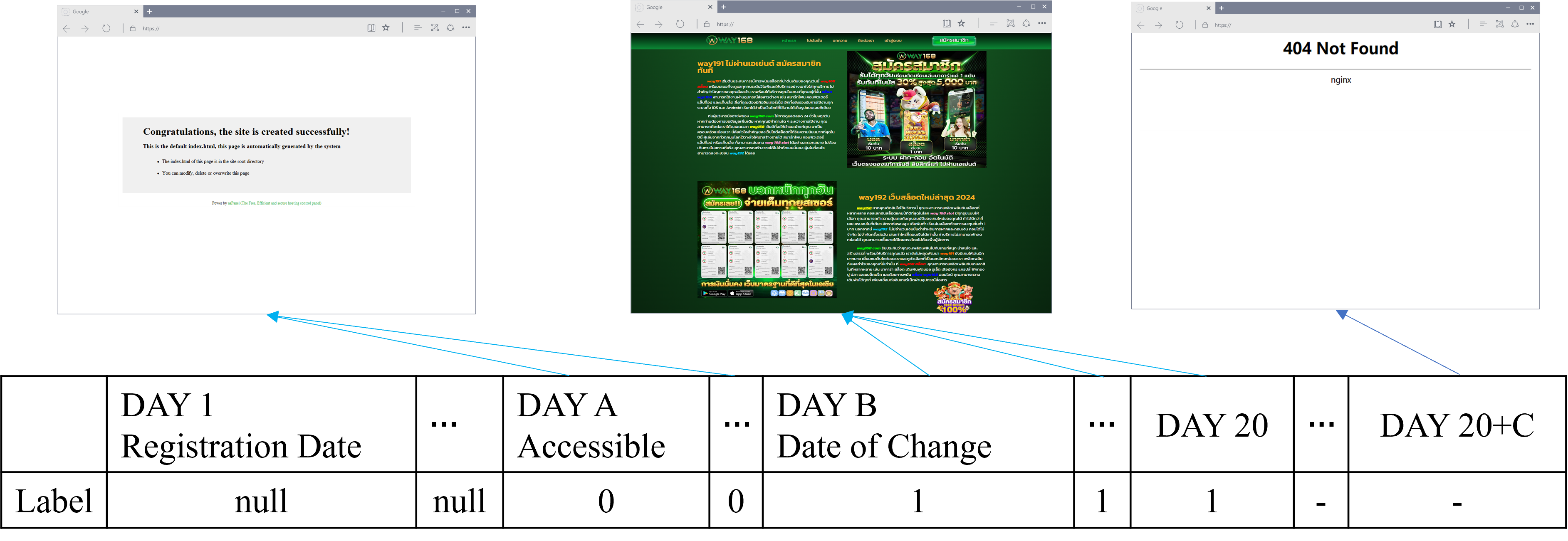}
\caption{The second part of the annotation. This part of the domain names are annotated for 20 days.}\label{fig 8}
\end{figure}

Our labels only indicate the nature of the site content at the moment of data capture. This means a domain may have different labels within a single detection period. For example, a domain might masquerade as a normal site or be under construction during the first \textit{n} days, and only on \textit{Day n+1} be utilized for pornographic and gambling purposes. Thus, our labeled data are divided into two parts: the first 2,000 domains are annotated based on site content from a randomly chosen day; the second part involves manually selected 100 domains registered on August 10, 2024, each annotated daily throughout the 20-day detection period, with these domains exhibiting real label changes during the detection cycle(as shown in Fig. \ref{fig 8}).

\subsection{Highly Adaptive Classification of PGDN}\label{subsec2 5}

\subsubsection{Data Augmentation}\label{subsubsec2 5 1}

Models trained and tested on ideal data often perform poorly when applied to real-world scenarios; thus, we opt to use real-world scraped data for training. To meet the recognition requirements of real data, we implement additional enhancements. First, regarding textual features, we observe a high degree of homogeneity despite the large number of domain names collected, making blind annotation for larger datasets unproductive. Some pornographic and gambling websites use human-recognizable obfuscated or slightly disordered natural language text to disguise their page titles and evade detection. Second, for numerical features, missing data poses a major challenge due to the diversity and multiplicity of these features. We enhance the data using strategies outlined in Table \ref{tab 5}. (1) \textbf{Discarding:} Select 400 records with complete textual features and remove their textual features. (2) \textbf{Randomly Zeroing:} Randomize zeroing of up to five out of 19 numerical features for 1,500 records. (3) \textbf{Random Replacement:} Randomly replace text characters for 100 records with textual features. Table \ref{tab 6} shows some examples of our data augmentation method. Ultimately, we obtain 4,000 records.

\begin{table}[h]
\caption{\centering Data augmentation methods and their proportions.}\label{tab 5}%
\begin{tabular}{@{\hspace{10pt}}p{6cm}@{\hspace{10pt}} p{2cm}@{\hspace{10pt}}}
\toprule
\textbf{Method}                         & \textbf{Proportion} \\ 
\midrule
Discarding Text Features                & 20\%                \\ 
Randomly Zeroing of Numerical Feature   & 75\%                \\ 
Random Replacement of Text Letters       & 5\%                 \\ 
\botrule
\end{tabular}
\end{table}

\begin{table}[ht]
    \centering
    \caption{\centering Data augmentation methods and their examples.}\label{tab 6}
    \begin{tabular}{@{}p{2.5cm} p{2.5cm} p{1cm} p{1cm} p{1cm} p{1cm}@{}}
        \toprule
        \textbf{Method} & \textbf{Textual Feature} & \multicolumn{4}{c}{\textbf{Numerical Features}} \\ \midrule
       Original & ``Porn" & 1 & 2 & ... & 3 \\
        Method1 & null & 1 & 2 & ... & 3 \\
        Method2 Result1& ``Porn" & 0 & 2 & ... & 3 \\
        Method2 Result2& ``Porn" & 1 & 0 & ... & 3 \\
        Method2 Result3& ``Porn" & 1 & 1 & ... & 0 \\
        Method3 & ``P0*n" & 1 & 2 & ... & 3 \\ \bottomrule
    \end{tabular}
    
\end{table}

\subsubsection{Our Two-level Classification Schema For PGDN}\label{subsubsec2 5 2}

Currently, a dataset of 4,000 entries is obtained for training and validation, each comprising 19 numerical features and one textual feature (i.e., the \textit{title}). Next, we employ our proposed two-level classification schema for PGDN, which integrates the CoSENT model, MLP model, and traditional classification algorithms to establish a PGDN classifier.

Initially, we utilize natural language processing techniques to extract features from the \textit{title}. Historically, feature extraction in text relied on manually constructed features and simple text statistics like word frequency. Subsequently, word embedding techniques, most notably the \textit{word2vec} algorithm \cite{bib9}, gained prominence. With the advent of deep learning, context-based embedding techniques emerged, and the BERT model, based on the Transformer architecture, achieved remarkable success. Sentence-BERT (SBERT)\cite{bib10}, proposed by Reimers and Gurevych, is more suitable for sentence-level tasks than BERT. In 2024, Huang et al. introduced CoSENT, which replaces SBERT's Cross-entropy Loss with Cosine Similarity Loss (see Equation \ref{eq 3} and Equation \ref{eq 4}) \cite{bib11,bib12}, yielding better text embedding performance. In this study, we use CoSENT to embed the textual feature to obtain a 384-dimensional vector.

\begin{equation}
\operatorname{Loss}_{\operatorname{SBERT}}=max (||s_ {a}-s_ {p}||-||s_ {a}-s_ {n}||+epsilon,0)
\label{eq 3}
\end{equation}

\begin{equation}
\operatorname{Loss}_{\operatorname{CoSENT}}=\log \left(1+\sum_{(i, j) \in \Omega_{\text {pos }},(k, l) \in \Omega_{\text {neg }}} e^{\lambda\left(\cos \left(u_{k}, u_{l}\right)-\cos \left(u_{i}-u_{j}\right)\right)}\right)
\label{eq 4}
\end{equation}

\begin{figure}[h]
\centering
\includegraphics[width=0.8\textwidth]{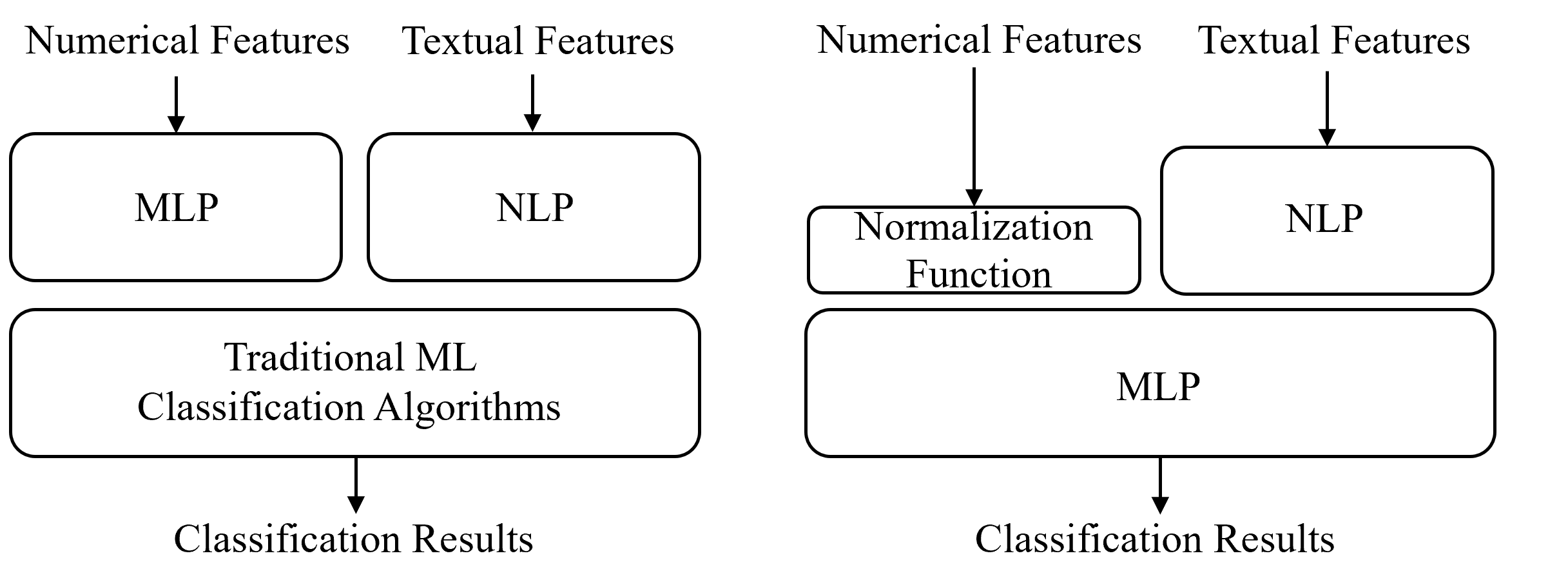}
\caption{\centering Potential combinations of NLP and MLP for domain classification.}\label{fig 9}
\end{figure}

Then, we consider incorporating the Multilayer Perceptron (MLP). The Multilayer Perceptron, also known as the Artificial Neural Network (ANN), is a type of feedforward neural network \cite{bib13}. In this paper, we will use it to build a classifier. Fig. \ref{fig 9} illustrates potential combinations of NLP and MLP for domain classification. The architecture shown on the left side of Fig. \ref{fig 9} has been used for domain classification \cite{bib14}, which segregates numerical and textual features, embeds them separately with MLP and NLP models, and finally applies an independent machine learning classification algorithm. However, our experiments show poor classification results when simply inputting NLP output into a machine learning algorithm. Moreover, for numerical features, machine learning generally requires no additional processing. Our proposed architecture (shown on the right side of Fig. \ref{fig 9}) emphasizes the role of MLP, where normalized numerical features and embedded textual feature vectors are fed into the MLP, directly outputting binary classification results through fully connected layers. However, this strategy of mixing feature vectors with no practical meaning and numerical features with practical meaning is illogical.

\begin{figure}[h]
\centering
\includegraphics[width=0.75\textwidth]{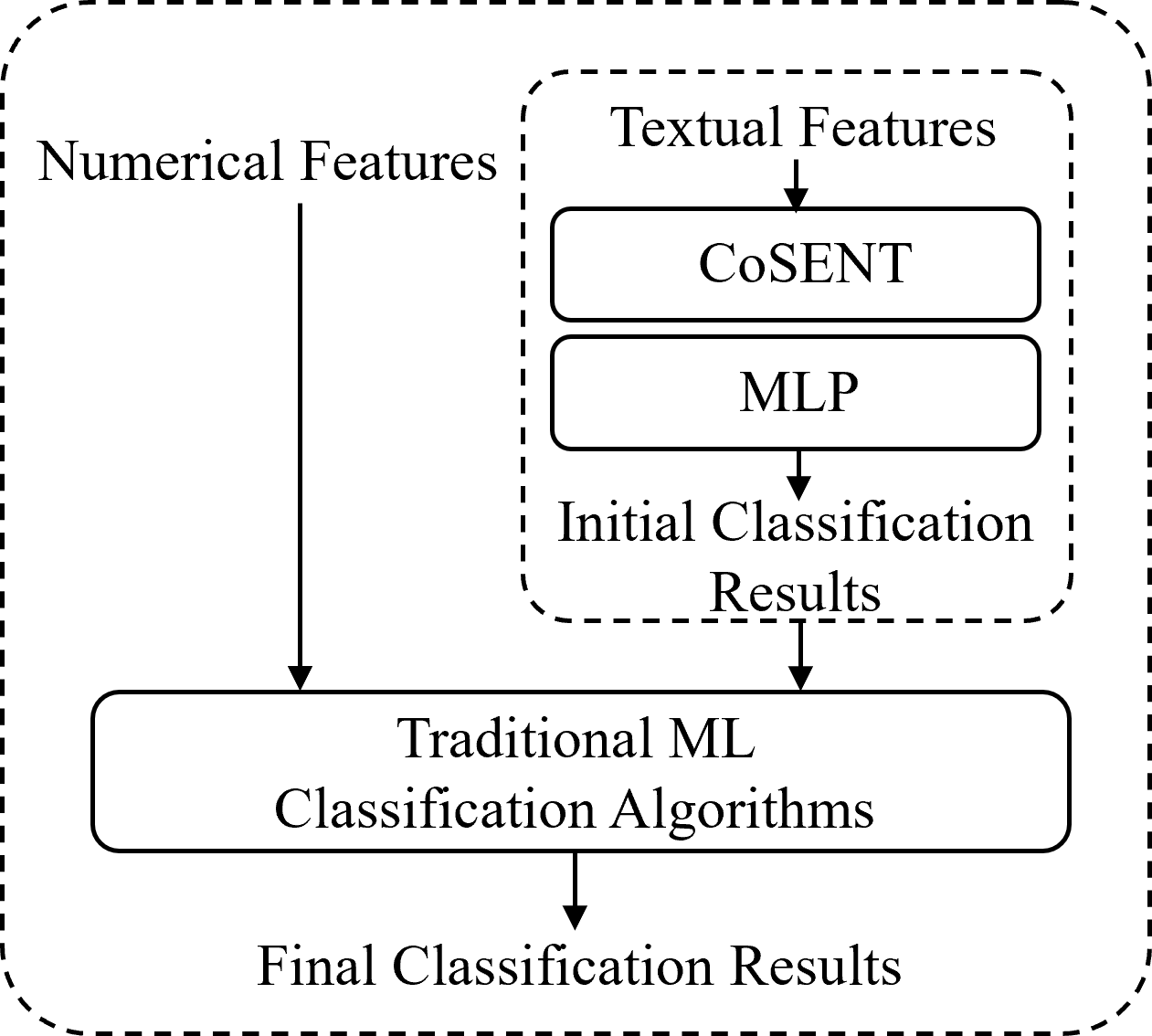}
\caption{\centering Our two-level classification schema}\label{fig 10}
\end{figure}

\begin{figure}[h]
\centering
\includegraphics[width=0.9\textwidth]{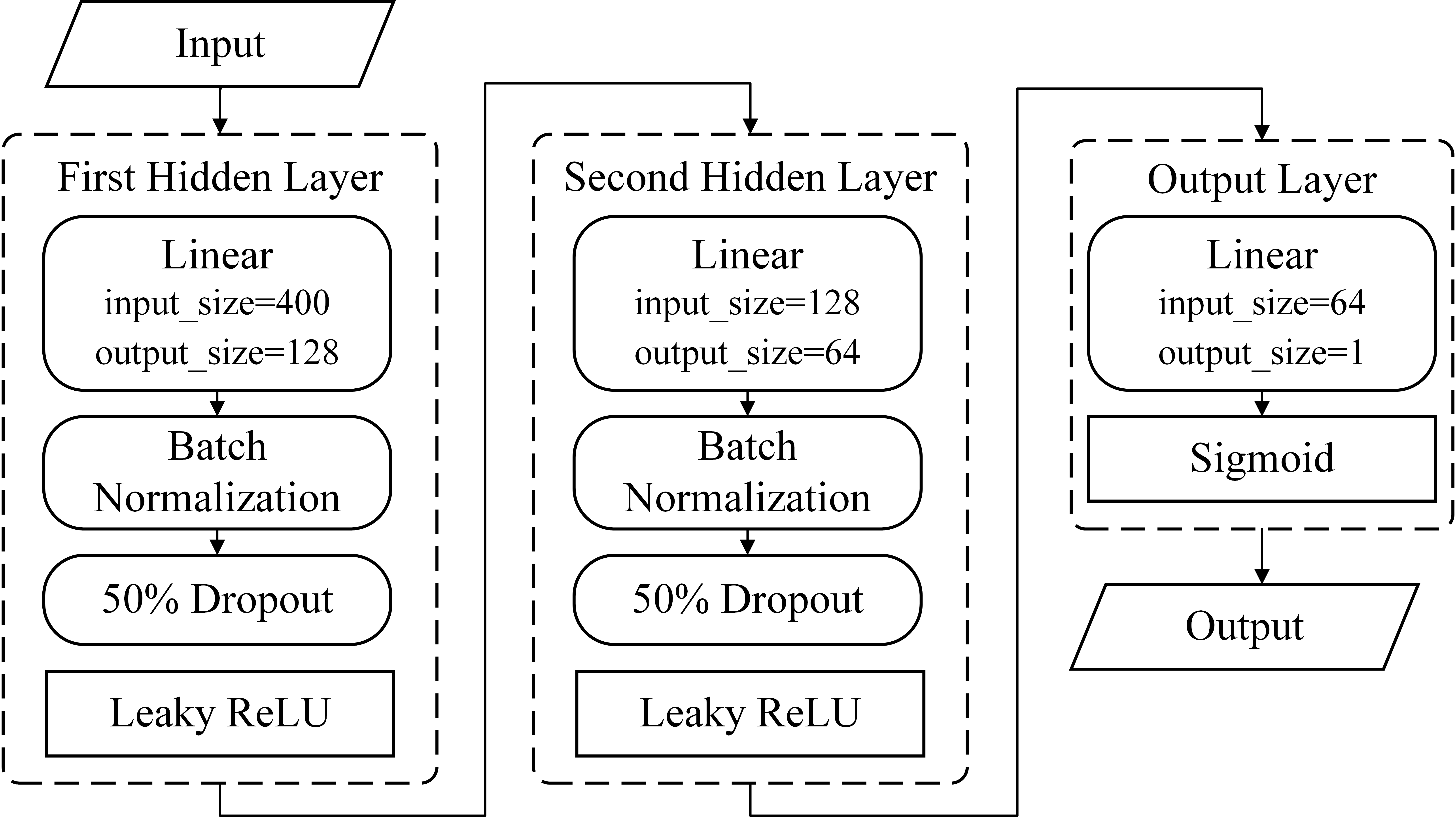}
\caption{\centering Model parameters of our MLP classifier.}\label{fig 11}
\end{figure}

Inspired by the above two architectures, we developed our two-level classification schema (as shown in Fig. \ref{fig 10}). Initially, CoSENT NLP model outputs are connected to the MLP classifier for the preliminary classification of textual features. Fig. \ref{fig 11} details the parameters of the MLP. As illustrated, MLP receives CoSENT outputs and performs the first classification via its two hidden layers and an output layer. Subsequently, all 19 numerical features, along with the binary initial classification result (0 or 1), are input into a separate machine learning classifier for the final classification. Classical machine learning classifiers can be flexibly adapted to real conditions. Additionally, our two-level classification schema is better suited to handle scenarios with randomly missing features. In the absence of textual features, the first classification is bypassed, and a 0 is output, ensuring that missing text features do not significantly affect classification outcomes, thereby achieving robust PGDN recognition.

\section{Result}\label{sec3}

\subsection{Evaluation}\label{subsec3 1}

In classification models, it is imperative to delineate between positive and negative data. In this paper, we define PGDN as positive samples, while others (let's call them benign domains for now) as negative samples. True positives (TP) represent correctly predicted PGDN, false positives (FP) indicate incorrectly predicted such domains, true negatives (TN) are rightly predicted benign domains, and false negatives (FN) represent erroneously predicted benign domains. We employ accuracy, precision, recall, and F1-score as evaluation parameters, with Equation \ref{eq 5},\ref{eq 6},\ref{eq 7},\ref{eq 8} depicted below. Accuracy measures the ratio of correctly classified predictions to the total number of predictions; precision quantifies the ratio of true positive samples to all predicted positive results; recall indicates the proportion of actual positive samples that are correctly predicted; and the F1-score is the weighted harmonic mean of precision and recall, taking both false positives and false negatives into account.

\begin{equation}
{ Accuracy }=\frac{T P+T N}{T P+F P+T N+F N}
\label{eq 5}
\end{equation}

\begin{equation}
{ Precision }=\frac{T P}{T P+F P}
\label{eq 6}
\end{equation}

\begin{equation}
{ Recall }=\frac{T P}{T P+F N}
\label{eq 7}
\end{equation}

\begin{equation}
F 1=\frac{2 * { Precision } * { Recall }}{{ Precision }+ { Recall }}
\label{eq 8}
\end{equation}

Our testing can be divided into three stages. Initially, we assess the classification effectiveness of SVM (Support Vector Machine), DT (Decision Tree), and RF (Random Forest) on numerical features, which is the most common method in previous research. Next, we incorporate textual features and conduct one set of experiments using the schema illustrated in Fig. \ref{fig 9}, which integrates NLP and MLP. Finally, we evaluate the performance of the two-level classification schema, adopting the same classical classification algorithms for comparative transparency in results. To mitigate the effects of randomness, we average the results over five tests. The detailed results are shown in Table \ref{tab 7}.

\begin{table}[h]
\caption{\centering Evaluation results for different models.}\label{tab 7}
\begin{tabular*}{0.98\textwidth}{@{\extracolsep\fill}lccccccc}
\toprule
\multirow{2}{*}{\textbf{Method}}& \multirow{2}{*}{\textbf{Evaluation}} & \multicolumn{5}{c}{\textbf {Number of Experiments}} & \multirow{2}{*}{\textbf{Average}} \\
\cmidrule(lr){3-7}
& & 1 & 2 & 3 & 4 & 5 \\
\midrule
\multirow{4}{*}{SVM (No Title)} & Accuracy & 0.8538 & 0.8530 & 0.8701 & 0.8600 & 0.8608 & 0.8595 \\
& Precision & 0.8110 & 0.8310 & 0.8272 & 0.8177 & 0.8256 & 0.8225 \\
& Recall & 0.7567 & 0.7558 & 0.7578 & 0.7663 & 0.7568 & 0.7587 \\
& F1-score & 0.7829 & 0.7916 & 0.7910 & 0.7912 & 0.7897 & 0.7893 \\
\midrule
\multirow{4}{*}{DT (No Title)} & Accuracy & 0.9028 & 0.9168 & 0.9036 & 0.9005 & 0.9082 & 0.9064 \\
& Precision & 0.8815 & 0.8527 & 0.8575 & 0.8495 & 0.8394 & 0.8561 \\
& Recall & 0.8322 & 0.9065 & 0.8689 & 0.8822 & 0.8876 & 0.8755 \\
& F1-score & 0.8562 & 0.8788 & 0.8631 & 0.8655 & 0.8628 & 0.8653 \\
\midrule
\multirow{4}{*}{RF (No Title)} & Accuracy & 0.9572 & 0.9603 & 0.9635 & 0.9627 & 0.9650 & 0.9617 \\
& Precision & 0.9606 & 0.9500 & 0.9732 & 0.9652 & 0.9539 & 0.9606 \\
& Recall & 0.9091 & 0.9246 & 0.9174 & 0.9265 & 0.9431 & 0.9241 \\
& F1-score & 0.9341 & 0.9371 & 0.9445 & 0.9455 & 0.9485 & 0.9419 \\
\midrule
\multirow{4}{*}{CoSENT + MLP} & Accuracy & 0.9386 & 0.9440 & 0.9362 & 0.9355 & 0.9401 & 0.9389 \\
& Precision & 0.9063 & 0.9260 & 0.9039 & 0.9020 & 0.9050 & 0.9086 \\
& Recall & 0.9204 & 0.9137 & 0.9159 & 0.9159 & 0.9270 & 0.9186 \\
& F1-score & 0.9133 & 0.9198 & 0.9099 & 0.9089 & 0.9158 & 0.9135 \\
\midrule
\multirow{4}{*}{\shortstack{CoSENT + MLP \\ + SVM}} & Accuracy & 0.8678 & 0.8655 & 0.8834 & 0.8802 & 0.8663 & 0.8726 \\
& Precision & 0.8363 & 0.8544 & 0.8354 & 0.8484 & 0.8212 & 0.8391 \\
& Recall & 0.7552 & 0.7617 & 0.7952 & 0.7904 & 0.7548 & 0.7715 \\
& F1-score & 0.7937 & 0.8054 & 0.8148 & 0.8184 & 0.7866 & 0.8038 \\
\midrule
\multirow{4}{*}{\shortstack{CoSENT + MLP \\+ DT}} & Accuracy & 0.9456 & 0.9479 & 0.9456 & 0.9479 & 0.9495 & 0.9473 \\
& Precision & 0.9361 & 0.9055 & 0.9426 & 0.9344 & 0.9373 & 0.9312 \\
& Recall & 0.9071 & 0.9379 & 0.8955 & 0.9110 & 0.9089 & 0.9121 \\
& F1-score & 0.9213 & 0.9215 & 0.9184 & 0.9225 & 0.9229 & 0.9213 \\
\midrule
\multirow{4}{*}{\textbf{\shortstack{CoSENT + MLP \\ + RF}}} & Accuracy & 0.9712 & 0.9759 & 0.9782 & 0.9712 & 0.9767 & \textbf{0.9746} \\
& Precision & 0.9667 & 0.9860 & 0.9837 & 0.9763 & 0.9812 & \textbf{0.9788} \\
& Recall & 0.9519 & 0.9441 & 0.9526 & 0.9385 & 0.9499 & \textbf{0.9474} \\
& F1-score & 0.9592 & 0.9646 & 0.9679 & 0.9570 & 0.9653 & \textbf{0.9628} \\
\bottomrule
\end{tabular*}
\end{table}

The test results suggest that using MLP to classify the title followed by integrating numerical features with Random Forest yields the best outcomes, with an accuracy of 0.9746, precision of 0.9788, recall of 0.9474, and an F1-score of 0.9628. While Random Forest performs best when solely utilizing numerical features, it is outperformed by the method incorporating textual features, which improved scores by 0.0129, 0.0182, 0.0233, and 0.0209 respectively, thus substantiating the value of textual features. Additionally, independently classifying the title before combining it with MLP for classification surpasses the collective vector approach, with respective improvements of 0.0357, 0.0702, 0.0288, and 0.0493, affirming the efficacy of our two-level classification schema.

Consequently, we select the CoSENT+MLP+RF approach to build our final classification model. Given that the correct prediction of pornographic and gambling domains (TP) is crucial, with benign domains potentially becoming PGDN in the future, precision is more valuable for reference. Our method excels in precision, achieving a rate of 0.9788.

\subsection{Analysis of Practical Application}\label{subsec3 2}

\begin{figure}[h]
\centering
\includegraphics[width=0.5\textwidth]{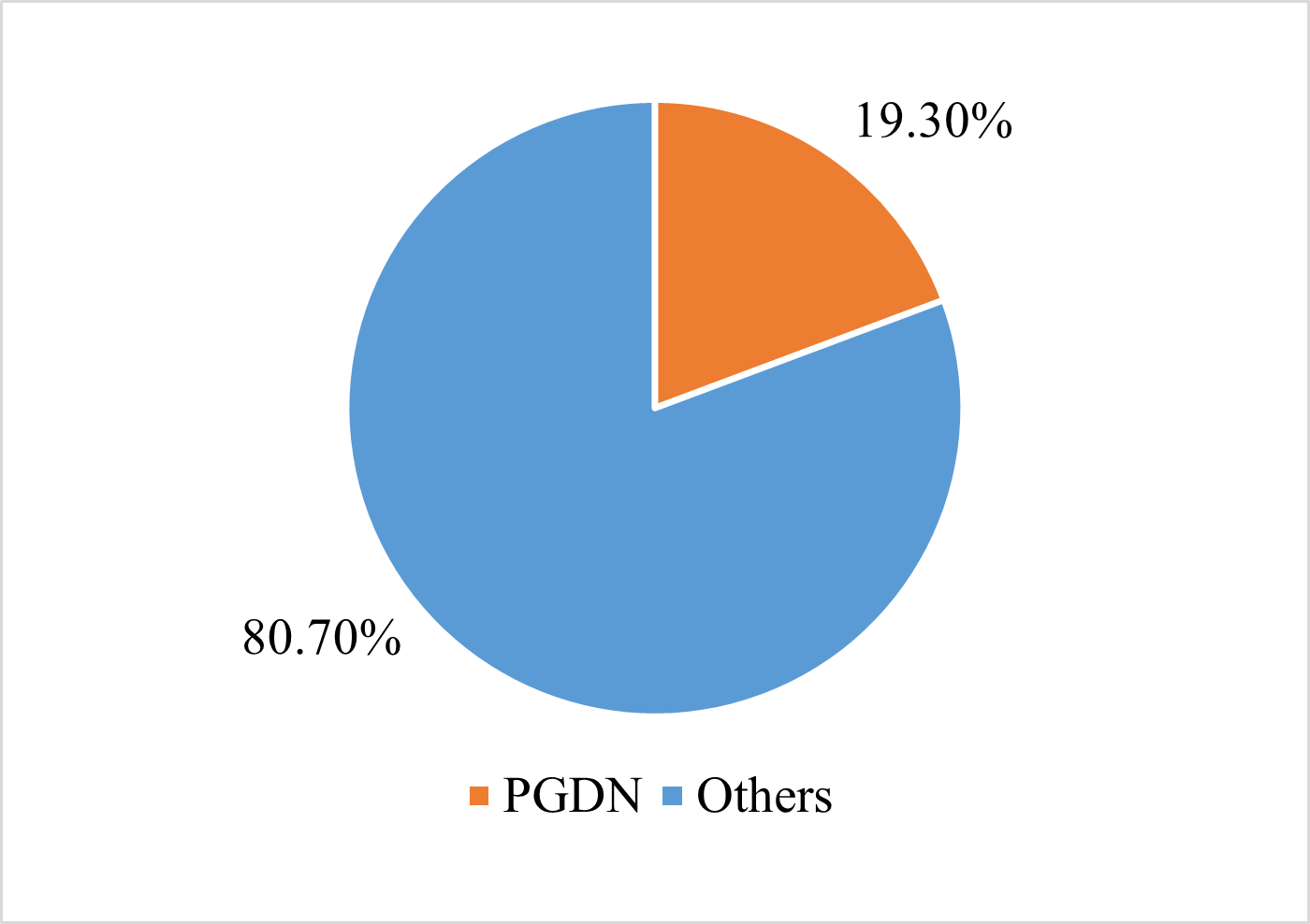}
\caption{\centering The proportion of PGDN in NRD2024 sampling data.}\label{fig 12}
\end{figure}

Firstly, we conduct an analysis to determine the proportion of NRD that constitutes PGDN. Utilizing the classifier optimized with CoSENT, MLP, and RF, as detailed in the previous section, we sample 10\% (152,236 entries) from the NRD2024 dataset for classification, identifying 29,381 PGDN, accounting for approximately 19.3\% (as shown in Fig. \ref{fig 12}).

\begin{figure}[h]
\centering
\includegraphics[width=0.9\textwidth]{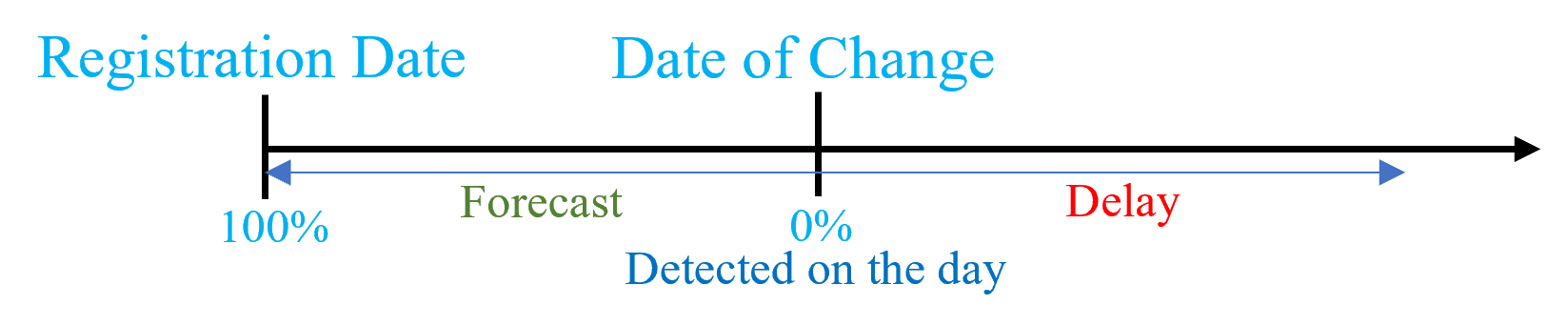}
\caption{The timeline of the detecting period. The time after registration is divided into ``Forecast" and ``Delay" by ``Date of Change”. In the forecast period, 100\% means complete prediction, and 0\% means just prediction.}\label{fig 13}
\end{figure}

Second, we evaluate the proposed classifier in real-world settings. The Real-PGDN method considers various dimensional features of domains, providing potential early prediction of PGDN. Hence, we manually annotate 100 domains from the NRD2024 dataset (Section \ref{subsec2 4}), which are later used for pornographic and gambling purposes. During the early period post-registration, these websites are displayed as under construction, parked pages, or disguised ``normal" pages. All 100 domain names are registered on August 10, 2024. Fig. \ref{fig 13} shows the timeline of the detecting period.

\begin{figure}[h]
\centering
\includegraphics[width=0.5\textwidth]{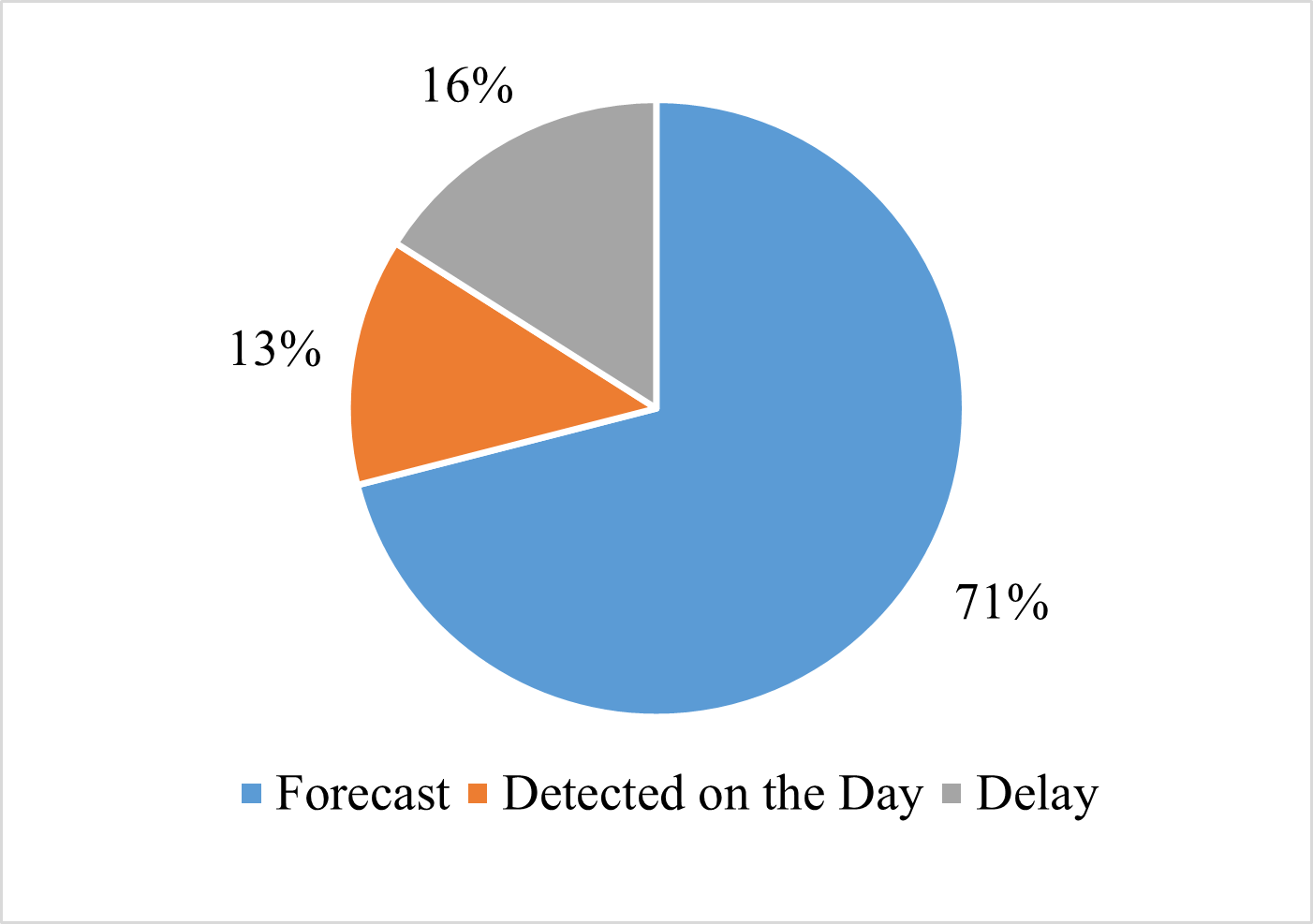}
\caption{\centering Classification statistics of sample PGDN data.}\label{fig 14}
\end{figure}

\begin{figure}[h]
\centering
\includegraphics[width=0.8\textwidth]{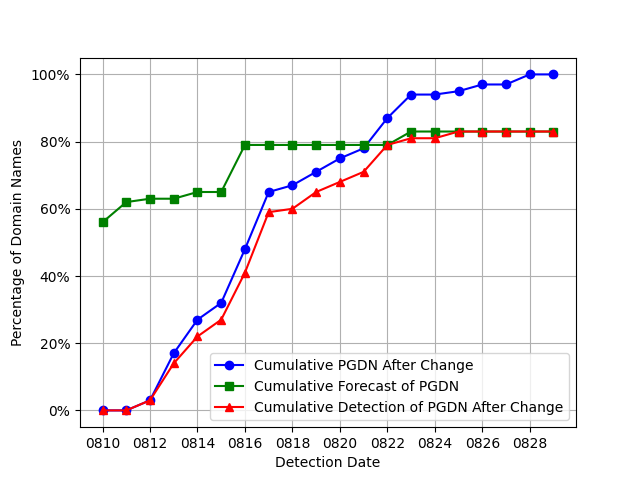}
\caption{\centering Statistics of PGDN changes, PGDN forecast, and PGDN detection.}\label{fig 15}
\end{figure}

\begin{figure}[h]
\centering
\includegraphics[width=0.8\textwidth]{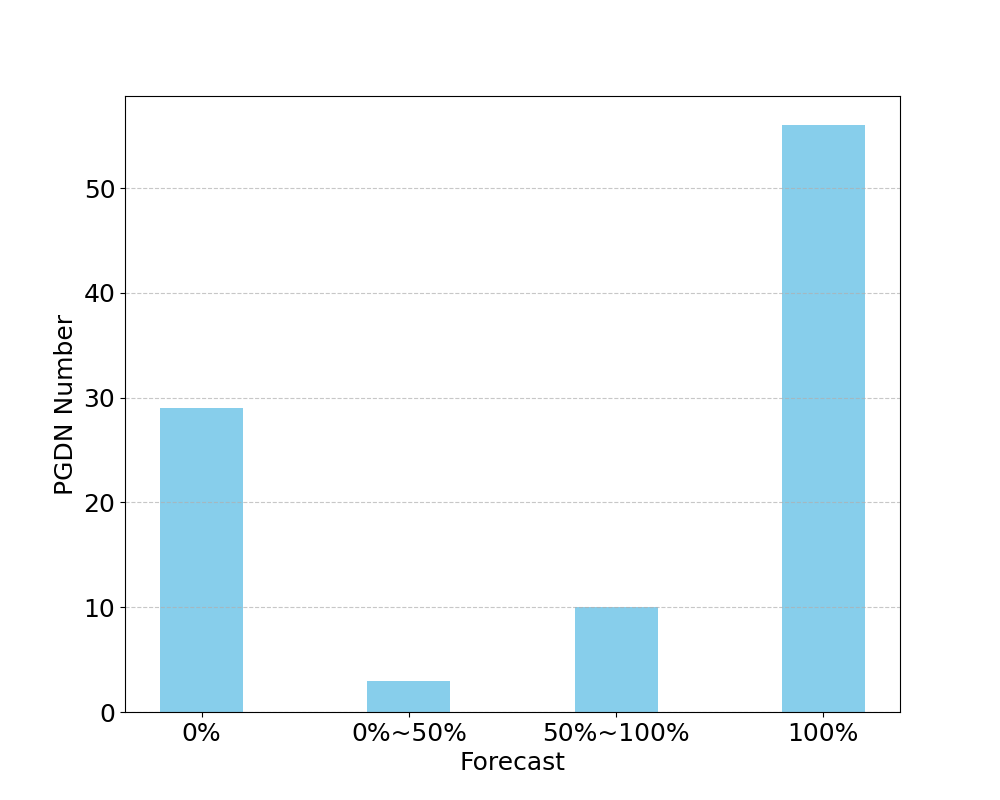}
\caption{\centering Statistics of the Forecast Rate of PGDN sampling data.}\label{fig 16}
\end{figure}

Our statistical results (as shown in Fig. \ref{fig 14}) show that approximately 71\% of PGDN are successfully forecasted, while 13\% are detected on the day their page ``turned bad". Since we do not possess data beyond the 20-day window for subsequent prediction, domains not detected on the given day are uniformly marked as delayed. Fig. \ref{fig 15} illustrates PGDN changes and detection trends. The blue curve shows the cumulative quantity of the PGDN with pages that have turned bad, the green curve presents the cumulative forecast of PGDN, and the red line illustrates the PGDN detection status after their pages have turned bad. We also analyze the lead time of PGDN predictions (as shown in Fig. \ref{fig 16}). Given that the actual date of usage for pornographic and gambling purposes varies among domains, we quantify the lead time using Equation \ref{eq 9}.

\begin{equation}
Forecast\ Rate =\frac{{ Days\ in\ advance }}{{ Date\ of\ change }-{ Registration\ Date }}
\label{eq 9}
\end{equation}

\section{Related Works}\label{sec4}

\subsection{Two-level Classifier}\label{subsec4 1}

A two-level classifier customarily comprises two hierarchical levels: the primary classifier executes preliminary categorization, whereas the secondary classifier conducts a more granular classification upon the initial outcomes. This methodology enhances the classifier's flexibility, accuracy, and adaptability. Some of the applications are as follows. Czyz et al. \cite{bib25} pioneered the integration of multiple classifiers in the domain of facial recognition. Zhang and Duin \cite{bib26} meticulously examined a variety of fusion strategies across different classifier combinations and conducted experiments across varying sample sizes. Song et al. \cite{bib27} utilized a two-level classifier to achieve more precise image categorization. In the field of credit risk assessment, Marqués et al. \cite{bib28} employed a two-level classifier. In recent research, Chakravarthy and Rajaguru \cite{bib24} successfully applied a two-level classifier in medical radiographic imaging for disease identification; Koshta and Singh \cite{bib29} utilized a two-level classification system, comprising KNN and CNN, for health prediction based on lung sounds. Predominantly, these studies employ secondary classifiers in image recognition. We have extended this dual-classification technique to the domain of domain name categorization, incorporating adjustments to the classifiers in accord with the distinctive characteristics of domain name data.

\subsection{Malicious Domain Classification}\label{subsec4 2}

Popular malicious domain classification methods can be divided into two categories: traditional machine learning methods and the burgeoning deep learning methods. Typically, domain classification segregates domains into malicious and benign categories. Malicious domains encompass phishing domains, malware distribution domains, illegal content hosting domains, spam domains, etc., often focusing on one or more specific subcategories. Davuth and Kim \cite{bib20} classified primarily botnet-related malicious domains using the Support Vector Machine (SVM). Cheng et al. \cite{bib30} compared the performance of the Logistic Regression, Random Forest, and AdaBoost algorithms for malicious domain classification based on WHOIS data. Silva et al. \cite{bib21} compared various machine learning algorithms, including Random Forest and Logistic Regression, to study malicious hosting domains systematically. In their research on newly registered malicious domains, Silveira et al. \cite{bib16} deployed machine learning algorithms, including optimized ones like XGBoost and LightGBM. Yang et al. \cite{bib22} proposed N-Trans, integrating N-gram methods with transformer models, yielding impressive results in malicious domain classification. Vranken and Alizadeh \cite{bib23} combined machine learning and deep learning for DGA domain detection. The research most analogous to ours is by Colhak et al. \cite{bib14}. They utilized two different NLP models for embedding webpage contents and titles and embedded numerical features through MLP before using a linear classifier for categorizing spam and phishing domains. While we adopted their idea of NLP model embeddings, we questioned their rationale for separately processing numerical features with MLP before continuing classification with classical algorithms. Ultimately, we converged upon an integrated approach using NLP, MLP, and classical classification algorithms, balancing accuracy and adaptability to real-world scenarios.

\subsection{Feature Analysis of NRD}\label{subsec4 3}

Commonly analyzed domain features include textual attributes, DNS records, site content, and WHOIS records. Many studies have utilized textual features of domain names \cite{bib6,bib7,bib8}, \cite{bib15}. Silveira et al. \cite{bib16} concentrated on using DNS information to detect newly registered malicious domains. Lauinger et al. \cite{bib17} employed WHOIS data in their study of domain expiration and re-registration. Alowaisheq et al. \cite{bib18} examined the Domain Take-down process based on DNS data and historical WHOIS features. Soska and Christin \cite{bib19} explored combining third-party service-provided site and content features. With the widespread adoption of the TLS protocol, research centered on digital certificate features has burgeoned. Sun et al. \cite{bib2} successfully classified gambling domains using textual and digital certificate features of websites. The study most closely related to this paper is Berenschot's research in 2024 \cite{bib3}, which classified malicious domains using a combination of features, including DNS, digital certificate, site characteristics, etc. This study synthesizes and integrates many features mentioned in the above research, introducing novel features and processing methods.

\section{Conclusion}\label{sec5}

This paper introduces an innovative Real-PGDN method that comprehensively achieves rapid and thorough detection of NRD information, effective extraction of PGDN features, and accurate recognition with fault tolerance for missing features in real-world scenarios. Our NRD2024 dataset consolidates extensive feature information from 1.5 million domains registered in 2024, facilitating future research. The model built upon our two-level classification schema exhibits a precision of 0.9788. Additionally, the case study demonstrates a forecast success rate exceeding 70\% for newly registered PGDN that are deferred for sensitive use. This advancement can significantly benefit Internet security by enabling quicker responses to potential threats.

\section*{Declarations}

\begin{itemize}
\item \textbf{Funding}: This work is supported by the Shandong Provincial Natural Science Foundation [Grant No. ZR2024QF138], the Scientific Research Innovation Fund of Harbin Institute of Technology [Grant No. IDGAZMZ00210335], and the Young Teacher Development Fund of Harbin Institute of Technology [Grant No. IDGA10002190].
\item \textbf{Conflict of interest}: All authors disclosed no relevant relationships.
\item \textbf{Ethics approval}: In this study, all data collection procedures strictly adhered to applicable laws and regulations. The data we gathered were sourced from publicly accessible information resources, negating the need for additional user consent.
\item \textbf{Consent for publication}: All authors have read and approved the final manuscript and consent to its publication.
\item \textbf{Data availability}: Data available within the article or its supplementary materials. The datasets generated or analyzed during this study are available in the Kaggle repository, \url{www.kaggle.com/datasets/hopehaowang/nrd2024}.
\item \textbf{Materials availability}: Materials available on request from the authors.
\item \textbf{Code availability}: Code available on request from the authors.
\item \textbf{Author contribution}: Hao Wang devised the project, developed the main conceptual ideas, and formulated all technical details. Yingshuo Wang made significant contributions to the experimental part of the paper. Junang Gan and Yanan Cheng contributed to the conceptual ideas. Jinshuai Zhang provided critical feedback on the manuscript.

\end{itemize}


\bibliography{realPGDNbib}

@inproceedings{bib1,
  title={Game of Registrars: An Empirical Analysis of $\{$Post-Expiration$\}$ Domain Name Takeovers},
  author={Lauinger, Tobias and Chaabane, Abdelberi and Buyukkayhan, Ahmet Salih and Onarlioglu, Kaan and Robertson, William},
  booktitle={26th USENIX Security Symposium (USENIX Security 17)},
  pages={865--880},
  year={2017}
}

@article{bib2,
  title={Gambling domain name recognition via certificate and textual analysis},
  author={Sun, GuoYing and Ye, Feng and Chai, Tingting and Zhang, Zhaoxin and Tong, Xiaojun and Prasad, Shitala},
  journal={The Computer Journal},
  volume={66},
  number={8},
  pages={1829--1839},
  year={2023},
  publisher={Oxford University Press}
}

@mastersthesis{bib3,
  title={Early Warning System for Newly Registered Malicious Domains: A Machine Learning and Certificate Transparency Approach},
  author={Berenschot, Luuk},
  year={2024},
  school={University of Twente}
}

@inproceedings{bib4,
  title={The analysis of the performance of RabbitMQ and ActiveMQ},
  author={Ionescu, Valeriu Manuel},
  booktitle={2015 14th RoEduNet International Conference-Networking in Education and Research (RoEduNet NER)},
  pages={132--137},
  year={2015},
  organization={IEEE}
}

@inproceedings{bib5,
  title={Modelling producer/consumer constraints},
  author={Simonis, Helmut and Cornelissens, Trijntje},
  booktitle={International Conference on Principles and Practice of Constraint Programming},
  pages={449--462},
  year={1995},
  organization={Springer}
}

@article{bib6,
  title={Exposure: A passive dns analysis service to detect and report malicious domains},
  author={Bilge, Leyla and Sen, Sevil and Balzarotti, Davide and Kirda, Engin and Kruegel, Christopher},
  journal={ACM Transactions on Information and System Security (TISSEC)},
  volume={16},
  number={4},
  pages={1--28},
  year={2014},
  publisher={ACM New York, NY, USA}
}

@inproceedings{bib7,
  title={COMAR: classification of compromised versus maliciously registered domains},
  author={Maroofi, Sourena and Korczy{\'n}ski, Maciej and Hesselman, Cristian and Ampeau, Benoit and Duda, Andrzej},
  booktitle={2020 IEEE European Symposium on Security and Privacy (EuroS\&P)},
  pages={607--623},
  year={2020},
  organization={IEEE}
}

@article{bib8,
  title={Malicious domain name detection based on extreme machine learning},
  author={Shi, Yong and Chen, Gong and Li, Juntao},
  journal={Neural Processing Letters},
  volume={48},
  number={3},
  pages={1347--1357},
  year={2018},
  publisher={Springer}
}

@article{bib9,
  title={Word2Vec},
  author={Church, Kenneth Ward},
  journal={Natural Language Engineering},
  volume={23},
  number={1},
  pages={155--162},
  year={2017},
  publisher={Cambridge University Press}
}

@article{bib10,
  title={Sentence-BERT: Sentence Embeddings using Siamese BERT-Networks},
  author={Reimers, N},
  journal={arXiv preprint arXiv:1908.10084},
  year={2019}
}

@article{bib11,
  title={Generalized cross entropy loss for training deep neural networks with noisy labels},
  author={Zhang, Zhilu and Sabuncu, Mert},
  journal={Advances in neural information processing systems},
  volume={31},
  year={2018}
}

@article{bib12,
  title={Learning similarity with cosine similarity ensemble},
  author={Xia, Peipei and Zhang, Li and Li, Fanzhang},
  journal={Information sciences},
  volume={307},
  pages={39--52},
  year={2015},
  publisher={Elsevier}
}

@article{bib13,
  title={Multilayer perceptron and neural networks},
  author={Popescu, Marius-Constantin and Balas, Valentina E and Perescu-Popescu, Liliana and Mastorakis, Nikos},
  journal={WSEAS Transactions on Circuits and Systems},
  volume={8},
  number={7},
  pages={579--588},
  year={2009},
  publisher={World Scientific and Engineering Academy and Society (WSEAS) Stevens Point~…}
}

@inproceedings{bib14,
  title={SecureReg: Combining NLP and MLP for Enhanced Detection of Malicious Domain Name Registrations},
  author={{\c{C}}olhak, Furkan and Ecevit, Mert {\.I}lhan and Da{\u{g}}, Hasan and Creutzburg, Reiner},
  booktitle={2024 International Conference on Electrical, Computer and Energy Technologies (ICECET},
  pages={1--6},
  year={2024},
  organization={IEEE}
}

@article{bib15,
  title={An ensemble approach for algorithmically generated domain name detection using statistical and lexical analysis},
  author={Anand, P Mohan and Kumar, T Gireesh and Charan, PV Sai},
  journal={Procedia Computer Science},
  volume={171},
  pages={1129--1136},
  year={2020},
  publisher={Elsevier}
}

@inproceedings{bib16,
  title={Detection of newly registered malicious domains through passive DNS},
  author={Silveira, Marcos Rog{\'e}rio and da Silva, Leandro Marcos and Cansian, Adriano Mauro and Kobayashi, Hugo Koji},
  booktitle={2021 IEEE International Conference on Big Data (Big Data)},
  pages={3360--3369},
  year={2021},
  organization={Ieee}
}

@inproceedings{bib17,
  title={WHOIS Lost in Translation: (Mis) Understanding Domain Name Expiration and Re-Registration},
  author={Lauinger, Tobias and Onarlioglu, Kaan and Chaabane, Abdelberi and Robertson, William and Kirda, Engin},
  booktitle={Proceedings of the 2016 Internet Measurement Conference},
  pages={247--253},
  year={2016}
}

@inproceedings{bib18,
  title={Cracking wall of confinement: Understanding and analyzing malicious domain takedowns},
  author={Alowaisheq, Eihal},
  booktitle={The Network and Distributed System Security Symposium (NDSS)},
  year={2019}
}

@inproceedings{bib19,
  title={Automatically detecting vulnerable websites before they turn malicious},
  author={Soska, Kyle and Christin, Nicolas},
  booktitle={23rd USENIX Security Symposium (USENIX Security 14)},
  pages={625--640},
  year={2014}
}

@article{bib20,
  title={Classification of malicious domain names using support vector machine and bi-gram method},
  author={Davuth, Nhauo and Kim, Sung-Ryul},
  journal={International Journal of Security and Its Applications},
  volume={7},
  number={1},
  pages={51--58},
  year={2013}
}

@inproceedings{bib21,
  title={Compromised or $\{$Attacker-Owned$\}$: A large scale classification and study of hosting domains of malicious $\{$URLs$\}$},
  author={De Silva, Ravindu and Nabeel, Mohamed and Elvitigala, Charith and Khalil, Issa and Yu, Ting and Keppitiyagama, Chamath},
  booktitle={30th USENIX security symposium (USENIX security 21)},
  pages={3721--3738},
  year={2021}
}

@article{bib22,
  title={N-trans: parallel detection algorithm for DGA domain names},
  author={Yang, Cheng and Lu, Tianliang and Yan, Shangyi and Zhang, Jianling and Yu, Xingzhan},
  journal={Future Internet},
  volume={14},
  number={7},
  pages={209},
  year={2022},
  publisher={MDPI}
}

@article{bib23,
  title={Detection of DGA-generated domain names with TF-IDF},
  author={Vranken, Harald and Alizadeh, Hassan},
  journal={Electronics},
  volume={11},
  number={3},
  pages={414},
  year={2022},
  publisher={MDPI}
}

@article{bib24,
  title={Performance analysis of ensemble classifiers and a two-level classifier in the classification of severity in digital mammograms},
  author={Sannasi Chakravarthy, SR and Rajaguru, Harikumar},
  journal={Soft Computing},
  volume={26},
  number={22},
  pages={12741--12760},
  year={2022},
  publisher={Springer}
}

@article{bib25,
  title={Multiple classifier combination for face-based identity verification},
  author={Czyz, Jacek and Kittler, Josef and Vandendorpe, Luc},
  journal={Pattern recognition},
  volume={37},
  number={7},
  pages={1459--1469},
  year={2004},
  publisher={Elsevier}
}

@article{bib26,
  title={An experimental study of one-and two-level classifier fusion for different sample sizes},
  author={Zhang, Chun-Xia and Duin, Robert PW},
  journal={Pattern Recognition Letters},
  volume={32},
  number={14},
  pages={1756--1767},
  year={2011},
  publisher={Elsevier}
}

@article{bib27,
  title={Two-level hierarchical feature learning for image classification},
  author={Song, Guang-hui and Jin, Xiao-gang and Chen, Gen-lang and Nie, Yan},
  journal={Frontiers of Information Technology \& Electronic Engineering},
  volume={17},
  number={9},
  pages={897--906},
  year={2016},
  publisher={Springer}
}

@article{bib28,
  title={Two-level classifier ensembles for credit risk assessment},
  author={Marqu{\'e}s, AI and Garc{\'\i}a, Vicente and S{\'a}nchez, Jos{\'e} Salvador},
  journal={Expert Systems with Applications},
  volume={39},
  number={12},
  pages={10916--10922},
  year={2012},
  publisher={Elsevier}
}

@inproceedings{bib29,
  title={A Two-Level Classifier for Prediction of Healthy and Unhealthy Lung Sounds Using Machine Learning and Convolutional Neural Network},
  author={Koshta, Vaibhav and Singh, Bikesh Kumar},
  booktitle={International Conference on Biomedical Engineering Science and Technology},
  pages={154--168},
  year={2023},
  organization={Springer}
}

@article{bib30,
  title={Detecting malicious domain names with abnormal whois records using feature-based rules},
  author={Cheng, Yanan and Chai, Tingting and Zhang, Zhaoxin and Lu, Keyu and Du, Yuejin},
  journal={The Computer Journal},
  volume={65},
  number={9},
  pages={2262--2275},
  year={2022},
  publisher={Oxford University Press}
}

\end{document}